\documentclass[11pt]{article}

\pdfoutput=1

\usepackage{booktabs} 

\usepackage[normalem]{ulem}

\usepackage{epsfig}
\usepackage{cite}
\usepackage{amsmath}
\usepackage[footnotesize]{caption}
\usepackage{slashed}
\usepackage{xcolor}
\usepackage[utf8]{inputenc}
\usepackage{tikz}
\usepackage{tikz-feynman}
\usetikzlibrary{shapes.geometric}
\usetikzlibrary{arrows.meta,arrows}
\usetikzlibrary{quotes,angles}

\usepackage{multicol}
\usepackage{multirow}
\usepackage{array}

\usepackage{cancel}

\numberwithin{equation}{section}

\newcommand{\be}{\begin{equation}}
\newcommand{\ee}{\end{equation}}
\newcommand{\beq}{\begin{eqnarray}}
\newcommand{\eeq}{\end{eqnarray}}

\usepackage{comment}

\begin{document}

\title{Impact of electroweak group representation in models for $B$ and $g-2$ anomalies from Dark Loops}

\date{\today}
\author{
Rodrigo Capucha$^{1\,}$\footnote{E-mail:
\texttt{rscapucha@fc.ul.pt}} ,
Da Huang$^{2,3\,}$\footnote{E-mail:
\texttt{dahuang@bao.ac.cn}} ,
Tomás Lopes$^{4\,}$\footnote{E-mail:
\texttt{tomasclopes@tecnico.ulisboa.pt}} ,
Rui Santos$^{1,5\,}$\footnote{E-mail:
  \texttt{rasantos@fc.ul.pt}} 
\\[5mm]
{\small\it $^1$Centro de F\'{\i}sica Te\'{o}rica e Computacional,
    Faculdade de Ci\^{e}ncias,} \\
{\small \it    Universidade de Lisboa, Campo Grande, Edif\'{\i}cio C8
  1749-016 Lisboa, Portugal} \\[3mm]
{\small\it
$^2$National Astronomical Observatories, Chinese Academy of Sciences, Beijing, 100012, China} \\[3mm]
{\small\it $^3$School of Fundamental Physics and Mathematical Sciences,} \\
{\small \it   Hanzhou Institute for Advanced Study, UCAS, Hanzhou 310024, China} \\[3mm]
{\small\it $^4$ CFTP, Departamento de F\'{\i}sica,} \\
{\small\it Instituto Superior T\'{e}cnico, Universidade de Lisboa,} \\
{\small\it Avenida Rovisco Pais 1, 1049-001 Lisboa, Portugal} \\[3mm]
{\small\it
$^5$ISEL -
 Instituto Superior de Engenharia de Lisboa,} \\
{\small \it   Instituto Polit\'ecnico de Lisboa
 1959-007 Lisboa, Portugal} \\[3mm]
}

\maketitle

\begin{abstract}
\noindent
We discuss two models which are part of a class providing a common explanation for lepton flavor universality violation in $b \to s l^+ l^- $ decays, the dark matter (DM) problem and the muon $(g-2)$ anomaly. The 
$B$ meson decays and the muon $(g-2)$ anomalies are explained by additional one-loop diagrams with DM candidates. The models have one extra fermion field
and two extra scalar fields relative to the Standard Model (SM). The $SU(3)$ quantum numbers are fixed by the interaction with the SM fermions in a new Yukawa Lagrangian
that connects the dark and the visible sectors. We compare two models, one where the fermion is a singlet and the scalars are doublets under $SU(2)_L$ and another one where
the fermion is a doublet and the scalars are singlets under $SU(2)_L$. We conclude that both models can explain all new physics phenomena simultaneously, 
while satisfying all other flavor and DM constraints. However, there are crucial differences between how the DM constraints affect the two models leading 
to a noticeable difference in the allowed DM mass range. 
\end{abstract}

\thispagestyle{empty}
\vfill
\newpage
\setcounter{page}{1}

\section{Introduction}
\hspace{\parindent} 

One of the main problems  at the core of any extension of the Standard Model (SM) is the existence of dark matter (DM). Although it is not at all clear if DM will manifest itself as a particle, this
is certainly an avenue of research that is worth exploring. In fact, a hypothetical DM particle is able to explain all the experimental evidence gathered so far
(see~\cite{Bertone:2016nfn} for a review). However, there are no restrictions regarding the nature of the DM particle. Not only the allowed mass range is almost unconstrained, but also
its quantum numbers are unknown. Therefore, as long as the experimental results are in agreement with the proposed DM candidate in a given model, all possibilities are in principle possible.
It would be interesting to have a DM candidate that could also solve other discrepancies observed in other and apparently unrelated  experiments.     

There are other hints of new physics in the particle physics realm. Such is the case of the observed anomalies in the semileptonic $B$ meson decay rates, suggesting a violation of lepton flavor universality. 
The most recent measurements of the ratios of the exclusive branching fractions, $R(K^{(*)}) = {\cal B}(B\to K^{(*)}\mu^+\mu^-)/{\cal B}(B \to K^{(*)}e^+ e^-)$, 
are the ones obtained by the LHCb Collaboration~\cite{LHCb:2021trn,LHCb:2019hip,LHCb:2017avl}, yielding
\begin{eqnarray}
R(K)  = 0.846^{+0.042+0.013}_{-0.039-0.012} \,,\quad\quad q^2 \in [1.1,6] {\rm GeV}^2\,,
\end{eqnarray}
and
\begin{eqnarray}
R(K^*) = \left\{ \begin{array}{cc}
0.660^{+0.110}_{-0.070} \pm 0.024\,, & q^2 \in [0.045,1.1] {\rm GeV}^2\,,\\
0.685^{+0.113}_{-0.069} \pm 0.047\,, & q^2 \in [1.1,6] {\rm GeV}^2\, ,
\end{array}\right.
\end{eqnarray}
where $q^2$ is the dilepton mass squared in the processes. The SM predictions for these observables are~\cite{Hiller:2003js,Bordone:2016gaq}
\begin{eqnarray}
R(K) = 1.0004(8)\,, \quad\quad q^2 \in [1.1,6] {\rm GeV}^2\,,
\end{eqnarray}
and
\begin{eqnarray}
R(K^*) =\left\{ \begin{array}{cc}
0.920\pm 0.007\,, & q^2 \in [0.045,1.1] {\rm GeV}^2\,, \\
0.996 \pm 0.002\,, & q^2 \in [1.1,6] {\rm GeV}^2\,.
\end{array}
\right.
\end{eqnarray}
The Belle Collaboration has also measured these quantities~\cite{Belle:2019oag,BELLE:2019xld},  but with larger error bars when compared with the LHCb results.
It is important to note that these observables are clean probes of NP since the uncertainties  stemming from the hadronic matrix elements cancel out~\cite{Hiller:2003js} (both the theoretical and the 
experimental ones). The measurements of other observables in rare $B$ meson decays further support the existence of anomalies. These include differential branching ratios~\cite{LHCb:2014cxe, LHCb:2015wdu, Belle:2009zue} and angular distributions~\cite{CDF:2011tds, CMS:2015bcy, Belle:2016xuo, BaBar:2015wkg, LHCb:2015svh, Belle:2016fev, CMS:2017rzx, ATLAS:2018gqc} in the decays $B\to \phi \mu^+\mu^-$ and $B\to K^{(*)} \mu^+ \mu^-$, which also 
deviate from their SM predictions. These observables are all ultimately related with the $b\to s\mu^+\mu^-$ transition. Many proposals have been put forward to solve these discrepancies. Some of the most popular solutions are to introduce 
 a $Z^\prime$~\cite{Buras:2013qja,Gauld:2013qja, Altmannshofer:2019xda, Lebbal:2020sqb, Capdevila:2020rrl} or a leptoquark~\cite{Bauer:2015knc, Angelescu:2018tyl, Angelescu:2019eoh, Balaji:2019kwe, Crivellin:2019dwb, Saad:2020ucl, Fuentes-Martin:2020bnh} (see {\it e.g.}, Ref.~\cite{Capdevila:2017bsm} for a review), or new exotic particles which generate one-loop penguin and box diagrams~\cite{Gripaios:2015gra, Arnan:2016cpy, Arnan:2019uhr, Hu:2019ahp, Hu:2020yvs}.

Another very important and long-standing hint of NP is related to the anomalous magnetic moment of the muon, $(g-2)_\mu$~\cite{ParticleDataGroup:2018ovx,Gorringe:2015cma}. The most recent calculation
of this quantity in the framework of the SM~\cite{PhysRevLett.121.022003, RBC:2018dos}  shows a 4.2$\sigma$ discrepancy from the experimental measurement~\cite{PhysRevLett.126.141801, Muong-2:2006rrc}. 
Let us define $\Delta a_\mu$ as the difference between the experimentally measured value, $a_\mu^{\text{exp}}$, and the SM prediction, $a_\mu^{\text{SM}}$, 
\begin{equation}
	\Delta a_\mu = a_\mu^{\text{exp}} - a_\mu^{\text{SM}} \approx (251 \pm 59) \times 10^{-11} \, ,
\end{equation}
where the error is the combination of the theoretical and experimental uncertainties. Future experiments such as the ones planned for J-PARC~\cite{Saito:2012zz} and Fermilab~\cite{Muong-2:2015xgu}
aim at a large reduction in this experimental uncertainty.

In this paper we propose to solve the three problems described above simultaneously. Models that have addressed at least two of those problems have been proposed in the past. The DM problem has already been investigated in various models~\cite{Vicente:2018xbv} which also address the $B$ meson decay anomalies, such as {\it e.g.}, Refs.~\cite{AristizabalSierra:2015vqb, Belanger:2015nma, Altmannshofer:2016jzy, Celis:2016ayl, Cline:2017lvv, Ellis:2017nrp, Baek:2017sew, Fuyuto:2017sys, Cox:2017rgn, Falkowski:2018dsl, Darme:2018hqg, Singirala:2018mio, Baek:2018aru, Kamada:2018kmi, Guadagnoli:2020tlx} for $Z^\prime$ models, Refs.~\cite{deMedeirosVarzielas:2015lmh, Cline:2017aed, Hati:2018fzc, Choi:2018stw, Datta:2019bzu} for leptoquark models, and Refs.~\cite{Bhattacharya:2015xha, Kawamura:2017ecz, Cline:2017qqu, Cerdeno:2019vpd, Barman:2018jhz, Darme:2020hpo} for models with one-loop solutions.
In a previous work~\cite{PhysRevD.102.075009} a set of models were proposed by extending the work Ref.~\cite{Cerdeno:2019vpd}.  The model in question was built with the addition of three new fields to the SM,  
an $SU(3)_c$ coloured scalar which is also an $SU(2)_L$ singlet, $\Phi_3$, one $SU(2)_L$ singlet colourless scalar, $\Phi_2$, and one $SU(2)_L$ doublet vectorlike fermion, $\chi$, with $0, \pm1$ electric charge. In this work we will discuss
a new model where the scalars are $SU(2)_L$ doublets and the fermion is an $SU(2)_L$ singlet. The aim is to understand what is the role played by the group representations in providing a simultaneous solution to the 
three problems. While the Yukawa Lagrangian has a similar structure, the scalar potential is different in the two cases. More importantly, in this new model the scalars will couple to gauge bosons giving rise to the possibility
of a change in DM related observables.

The paper is organized as follows. In Sec.~\ref{sec:2}, we present the two models with the focus on the new one. In Sec.~\ref{sec:3} we discuss the flavor constraints and in Sec.~\ref{sec:4} we present the DM constraints on the model.
In Sec.~\ref{sec:5} we present and discuss our results. Finally, conclusions are given in Sec.~\ref{sec:6}.

\section{The models}
\hspace{\parindent} 
\label{sec:2}

In a previous work~\cite{PhysRevD.102.075009}, some of us have considered a model where three new fields were added to the SM,  one $SU(3)_c$ coloured scalar, $\Phi_3$, one colourless scalar, $\Phi_2$, and one vectorlike fermion, $\chi$, 
with an integer electric charge of 0 or $\pm 1$. In that work the scalars were $SU(2)_L$ singlets while the fermion was an $SU(2)_L$ doublet. That model was termed Model 5. We will now compare it to the scenario where the scalars are $SU(2)_L$ doublets and the fermion is an $SU(2)_L$ singlet. This model will be called Model 3 from now on. The complete set of quantum numbers is shown in Tables~\ref{table:1} and~\ref{table:2} for Models 3 and 5, respectively. 
\begin{table}[h!]
	\begin{center}
	\begin{tabular}{ |c|c|c|c| } 
		\hline
		& SU(3)$_c$ & SU(2)$_L$ & U(1)$_Y$ \\ \hline
		$\chi_R$ & 1 & 1 & -1 \\ 
		$\Phi_2$ & 1 & 2 & 1/2 \\
		$\Phi_3$ & 3 & 2 & 7/6 \\
		\hline
	\end{tabular}
	\end{center}
	\caption{SU(3)$_c$, SU(2)$_L$ and U(1)$_Y$ assignments for the newly introduced fields in Model 3.}
	\label{table:1}
\end{table}
\begin{table}[h!]
	\begin{center}
	\begin{tabular}{ |c|c|c|c| } 
		\hline
		& SU(3)$_c$ & SU(2)$_L$ & U(1)$_Y$ \\ \hline
		$\chi_R$ & 1 & 2 & -1/2 \\ 
		$\Phi_2$ & 1 & 1 & 0 \\
		$\Phi_3$ & 3 & 1 & 2/3 \\
		\hline
	\end{tabular}
	\end{center}
	\caption{SU(3)$_c$, SU(2)$_L$ and U(1)$_Y$ assignments for the newly introduced fields in Model 5.}
	\label{table:2}
\end{table}

A discrete $Z_2$ symmetry is imposed such that the SM fields are all even and the new fields are all odd under $Z_2$. The electric charges of the remaining fields can be determined from the following Yukawa Lagrangian
\begin{equation}
	\mathcal{L}^{\text{NP}}_{\text{Yuk}} = y_{Q_i} \overline{Q}_{Li} \Phi_3 \chi_R + y_{L_i} \overline{L}_{Li} \Phi_2 \chi_R + H.c. \, \, ,
	\label{eq:0}
\end{equation}
where $y_{Q_i}$ and $y_{L_i}$ are constants, $Q_{Li}$ and $L_{Li}$ the SM left-handed doublets for the quarks and leptons, respectively, and $\chi_R$ is the right-handed component of the new fermion, an $SU(2)_L$ singlet (doublet)
in Model 3 (Model 5). The scalar fields $\Phi_2$ and $\Phi_3$ are $SU(2)_L$ doublets (singlets) in Model 3 (Model 5). This new Yukawa Lagrangian connects the $Z_2$-odd dark sector with the $Z_2$-even SM and is necessary to explain the $B$ anomalies via one-loop diagrams.


The two sectors also communicate via the Higgs potential. In Model 3 where all scalar fields are $SU(2)_L$ doublets the potential can be written as (taking all parameters to be real) 
\begin{align}
	V =& -m_{11}^2  \Phi_1^\dagger \Phi_1 + m_{22}^2 \Phi_2^\dagger \Phi_2 +  m_{33}^2  \Phi_3^\dagger \Phi_3 + \lambda_1 (\Phi_1^\dagger \Phi_1)^2 + \lambda_2  (\Phi_2^\dagger \Phi_2)^2 - \lambda_{3} (\Phi_{3, a}^\dagger \Phi_{3, a})(\Phi_{3, b}^\dagger \Phi_{3, b}) \nonumber\\ 
	& + \lambda_{12} (\Phi_1^\dagger \Phi_1) (\Phi_2^\dagger \Phi_2) + \lambda_{13} (\Phi_1^\dagger \Phi_1) (\Phi_3^\dagger \Phi_3) + \lambda_{23} (\Phi_2^\dagger \Phi_2) (\Phi_3^\dagger \Phi_3) + \lambda_{5} \, [ \, (\Phi_1^\dagger \Phi_2)^2 + (\Phi_2^\dagger \Phi_1)^2 \, ] \nonumber\\ 
	& + \lambda_{12}' (\Phi_1^\dagger \Phi_2) (\Phi_2^\dagger \Phi_1) + \lambda_{13}' (\Phi_1^\dagger \Phi_3) (\Phi_3^\dagger \Phi_1) + \lambda_{23}' (\Phi_2^\dagger \Phi_3) (\Phi_3^\dagger \Phi_2)  \nonumber\\ 
	& + y_{13} (\Phi_1^T i \sigma_2 \Phi_3)^\dagger(\Phi_1^T i \sigma_2 \Phi_3) + y_{23} (\Phi_2^T i \sigma_2 \Phi_3)^\dagger(\Phi_2^T i \sigma_2 \Phi_3) \, , \label{eq:pot} \\ \nonumber  
\end{align}
with 
\begin{align}
	\Phi_1 = \begin{bmatrix}
			  	0 \\
			  	\frac{1}{\sqrt{2}} \, (v + h) 
			 \end{bmatrix} \, , \quad 
	\Phi_2 = \begin{bmatrix}
				\phi_{l}^+ \\
				\frac{1}{\sqrt{2}} \, (S + i A) 
			 \end{bmatrix} \, , \quad 
	\Phi_3 = \begin{bmatrix}
				\phi_{q}^{+5/3} \\
				\phi_{q}^{+2/3}
			 \end{bmatrix} \, , \\ \nonumber
\end{align}
in the unitary gauge. The $\Phi_1$ field is the SM Higgs doublet, $v$ its vacuum expectation value (VEV) and $h$ the SM Higgs field, with $v \approx 246$~GeV. Furthermore, $\sigma_2$ is the second Pauli matrix. We generally omit the $\Phi_3$ colour indices (a summation over colour is implied), except for the term proportional to $\lambda_3$, since the colour indices $a$ and $b$ may be different. Notice that the potential in Eq.~(\ref{eq:pot}) is the same as the one for the inert Two-Higgs-Doublet model (i2HDM) if we just consider the fields $\Phi_1$ and $\Phi_2$~\cite{2006}. The remaining terms include all the possibilities which are invariant under all symmetries when the $\Phi_3$ field is present which include terms of the type $y_{j3} (\Phi_j^T i \sigma_2 \Phi_3)^\dagger(\Phi_j^T i \sigma_2 \Phi_3)$~\cite{2022}. For the colourless doublets these terms are already present in the i2HDM potential.  

%

Since only the SM-Higgs doublet, $\Phi_1$, acquires a VEV, we have one minimization condition given by $m_{11}^2 = v^2 \lambda_1$, allowing to exchange one of the parameters by the VEV. The Higgs potential has therefore 15 independent (free) parameters. We have chosen as free input parameters of the potential all masses of the six physical Higgs bosons, 
and the quartic parameters $\lambda_2$, $\lambda_3$, $\lambda_{12}$,  
$\lambda_{13}$, $\lambda_{23}$, $\lambda_{23}'$, $y_{13}$ and $y_{23}$, together with the VEV that will be fixed by the W mass. 
Note that the couplings of the SM-like Higgs are exactly the same as in the SM.
As we are choosing the masses of the physical Higgs bosons to be free input parameters, the following parameters of the scalar potential were fixed:
\begin{align}
		\lambda_1 &= \frac{m_h^2}{2 v^2} \, , \quad m_{22}^2 = \frac{2 m_{\phi_{l}}^2 - v^2 \lambda_{12}}{2} \, , \quad m_{33}^2 = \frac{2 m_{\phi_{q}^{5/3}}^2 - v^2 y_{13} - v^2 \lambda_{13}}{2} \, , \\
		\quad \lambda_5 &= \frac{m_S^2 - m_A^2}{2 v^2} \, , \quad \lambda_{12}' = \frac{m_S^2 + m_A^2 - 2 m_{\phi_{l}}^2}{v^2} \, , \quad \lambda_{13}' = \frac{2 m_{\phi_{q}^{2/3}}^2 - 2 m_{\phi_{q}^{5/3}}^2 + v^2 y_{13} }{v^2} \, , 
		\label{eq:param} \\ \nonumber
\end{align}
where $m_S$ and $m_A$ are the masses of the neutral CP-even and CP-odd scalars $S$ and $A$, respectively, $m_{\phi_{l}}$ the mass of the charged scalar in the $\Phi_2$ doublet, and $m_{\phi_{q}^{5/3}}$ and $m_{\phi_{q}^{2/3}}$ the masses of the coloured scalars $\phi_{q}^{\pm 5/3}$ and $\phi_{q}^{\pm 2/3}$, respectively, from the $\Phi_3$ field. \\ 


Both models could in principle have the new fermion field as a DM candidate. However, as shown in~\cite{PhysRevD.102.075009} for Model 5  direct detection constraints exclude this possibility due to the tree-level Z mediation. The only way to avoid this limit would be to push the fermion mass to be  of ${\cal O}$(TeV) which in turn would make the loop contributions to $b \to s \mu^+ \mu^-$ and $\Delta a_\mu$ negligible and therefore the associated flavor anomalies would not be solved even for large Yukawa couplings. In Model 3 
the vectorlike fermion is charged and therefore cannot be the DM candidate. As such, the DM candidate can only come from the neutral components contained in the doublet scalar field $\Phi_2$, $S$ and $A$. In the previous study of Model 5~\cite{PhysRevD.102.075009}, the DM candidate was also chosen to be in $\Phi_2$, so in both studies the DM candidate comes from a scalar field. Although we chose $S$ to be the DM particle, assuming $m_S < m_A$, no differences were found in the final results when $A$ was chosen to be the DM candidate. 

The Dirac mass of $\chi$ is given by the term $m_\chi \overline{\chi}_L \chi_R + h.c.$. The Yukawa interaction in Eq.~(\ref{eq:0}) can be rewritten as follows
\begin{equation}
	\mathcal{L} = y_{di} (\overline{u}_{Lj} V_{ji} \chi_R^- \phi_q^{+5/3} + \overline{d}_{Li} \chi_R^- \phi_q^{+2/3}) + y_{Li} (\overline{\nu}_{Li} \chi_R^- \phi_l^{+} + \frac{\overline{e}_{Li}}{\sqrt{2}} \chi_R^- (S + i A)) + H.c. \, \, ,
\end{equation}
where $y_{di}$ is the new coupling when we write quarks in their mass eigenstates, and the matrix $V$ is the Cabibbo-Kobayashi-Maskawa (CKM) matrix. In order to suppress the strong flavor constraints on the first-generation of quarks and leptons and keep our analysis as simple as possible, we only take $y_b$, $y_s$ and $y_\mu$ to be nonzero.

Since we have introduced several new particles in our model, there can be corrections to the electroweak (EW) oblique parameters $S$, $T$ and $U$~\cite{PhysRevLett.65.964, PhysRevD.46.381}. We recall that in Model 5, the contribution to these parameters is zero. In this paper, we only consider the limits on the most relevant parameter, $T$.  The singlet vectorlike fermion, $\chi$, has a vanishing contribution to $T$, since the amplitude for the vacuum polarization diagram induced by this fermion at the one-loop level has a similar form to the one for the photon self-energy in QED (which is zero as the momentum transfer goes to zero). Thus, only the scalar fields can induce nonzero contributions to $T$. For this calculation, we followed~\cite{2008}, where a general expression for the oblique parameter $T$ is derived in the $SU(2)_L \times U(1)$ electroweak model with an arbitrary number of scalar doublets, with hypercharges $\pm 1/2$, and also an arbitrary number of scalar singlets. In Model 3, if we just consider the fields $\Phi_1$ and $\Phi_2$, this corresponds exactly to a 2HDM with a dark doublet~\cite{2006}, where the new physics (NP) contribution to $T$ is given as follows~\cite{2008}:
\begin{equation}
	T = \frac{g^2}{64 \pi^2 m_W^2 \alpha} [F(m_{\phi_l}^2, m_S^2) + F(m_{\phi_l}^2, m_A^2) - F(m_S^2, m_A^2)] ,
	\label{eq:T}
\end{equation}
where $m_W$ is the mass of the $W^\pm$ gauge boson, $\alpha$ is the fine-structure constant, $g$ is the $SU(2)_L$ coupling constant and the function $F(A, B)$ is defined as
\begin{equation}
	F(A, B) = \begin{cases}
		\frac{A + B}{2} - \frac{AB}{A - B} \ln \frac{A}{B}, & \text{if $A \neq B$}.\\
		0, & \text{ if A = B}.
	\end{cases} 
	\label{eq:FfromT}
\end{equation}

In a similar way, it can be shown that the contribution of the coloured scalar fields to $T$, $T^{c}$, is proportional to $F(m_{\phi_q^{5/3}}^2, m_{\phi_q^{2/3}}^2)$. Since we will always consider $m_{\phi_q^{5/3}} = m_{\phi_q^{2/3}}$ in our paper, this term will not contribute to $T$ and therefore the total NP contribution from Model 3 to the oblique parameter $T$ is given by Eq.~(\ref{eq:T}).
The most up-to-date~\cite{Zyla:2020zbs} value of this parameter is $T = 0.03 \pm 0.12$. This constraint will be applied at the end of our scan, with the requirement that every point in the allowed parameter space must be within the $2\sigma$ confidence interval experimentally observed.

\section{Flavor constraints}
\hspace{\parindent} 
\label{sec:3}

In this section we will discuss the flavor constraints. Not only have we to solve the discrepancies observed experimentally but we also have to make sure that the
observables that are in agreement with the SM predictions are not modified.
Since Model 3 and Model 5 have exactly the same NP contributions to the several flavor observables that are relevant, we will simply take the constraints and the analytic expressions used in our previous study of Ref.~\cite{PhysRevD.102.075009}. 

We start by the anomalous magnetic moment of the muon. Currently, the prediction of this quantity in the SM~\cite{PhysRevLett.121.022003} shows a 4.2$\sigma$ discrepancy from the experimental measurement~\cite{PhysRevLett.126.141801}. We define $\Delta a_\mu$ as the difference between the experimental measurement, $a_\mu^{\text{exp}}$, and the SM prediction, $a_\mu^{\text{SM}}$, with
\begin{equation}
	\Delta a_\mu = a_\mu^{\text{exp}} - a_\mu^{\text{SM}} \approx (251 \pm 59) \times 10^{-11} \, .
\end{equation}
In Model 3, like in Model 5, the leading-order (LO) contribution to this quantity is given by the one-loop diagrams enclosed by the fermion $\chi$ and the neutral scalars $S$ or $A$, with~\cite{2017}
\begin{equation}
	\Delta a_\mu = \frac{m_\mu^2 |y_\mu|^2}{16 \pi^2 m_\chi^2} (\tilde{F}_7 (x_S) + \tilde{F}_7 (x_A))
	\label{eq:delta_amu}
\end{equation}
and
\begin{equation}
	\tilde{F}_7 (x) = \frac{1 - 6x + 3x^2 + 2x^3 - 6x^2 \ln x}{12 (1-x)^4}, \quad \quad \quad x_i = m^2_i/m_\chi^2 .
\end{equation}

Regarding the anomalies in $B$ meson decays, they can be explained by using the effective Hamiltonian for $b \to s \mu^+ \mu^-$~\cite{Altmannshofer_2009, PhysRevD.86.034034}
\begin{equation}
	\mathcal{H}_{\text{eff}} = - \frac{4 G_F}{\sqrt{2}} V_{tb} V_{ts}^* (C_9^{\text{NP}} \mathcal{O}_9 + C_{10}^{\text{NP}} \mathcal{O}_{10}) ,
\end{equation}
with 
\begin{equation}
	\mathcal{O}_9 = \frac{\alpha}{4 \pi} (\overline{s} \gamma^\mu P_L b)(\overline{\mu} \gamma_\mu \mu), \quad 	\mathcal{O}_{10} = \frac{\alpha}{4 \pi} (\overline{s} \gamma^\mu P_L b)(\overline{\mu} \gamma_\mu \gamma^5 \mu).
\end{equation}
The main contribution to these operators comes from the box diagram shown in Figure~\ref{fig:box}, and the Wilson coefficients are given by~\cite{2017, PhysRevD.102.075009}
\begin{figure}[h!]
	\begin{center}
		\begin{tikzpicture}
			\begin{feynman}
				\vertex (a) {\(b\)};
				\vertex [right=of a] (b);
				\vertex [right=of b] (c);
				\vertex [above=0.05cm of c] {\(\chi^-\)};
				\vertex [right=of c] (d);
				\vertex [right=of d] (e) {\(\mu^-\)};
				\vertex [below=0.75cm of b] (b1);
				\vertex [left=0.05cm of b1] {\(\phi_q^{+2/3}\)};
				\vertex [below=0.75cm of d] (d1);
				\vertex [right=0.05cm of d1] {\(S/A\)};
				\vertex [below=0.75cm of b1] (b2);
				\vertex [below=0.75cm of d1] (d2);
				\vertex [left=of b2] (f) {\(\overline{s}\)};
				\vertex [right=of b2] (g);
				\vertex [below=0.05cm of g] {\(\chi^-\)};
				\vertex [right=of g] (h);
				\vertex [right=of h] (i) {\(\mu^+\)};
				\diagram* {
					(a) -- [] (b),
					(b) -- [] (c),
					(c) -- [] (d),
					(d) -- [] (e),
					(b) -- [scalar] (b1),
					(b1) -- [scalar] (b2),
					(b2) -- [] (f),
					(b2) -- [] (g),
					(g) -- [] (h),
					(d) -- [scalar] (d1),
					(d1) -- [scalar] (d2),
					(h) -- [] (i),
				};
			\end{feynman}
		\end{tikzpicture}
		\caption{One-loop box diagrams representing the dominant contribution to the operators $\mathcal{O}_9$ and $\mathcal{O}_{10}$ that were introduced to explain the flavor anomalies observed in $B$ meson decays.}
		\label{fig:box}
	\end{center}
\end{figure}
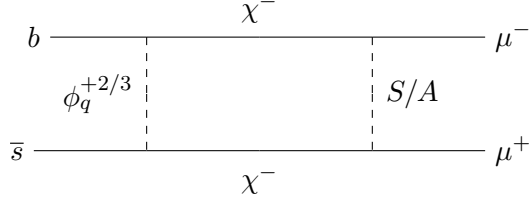 
\begin{equation}
	C_9^{\text{NP}} = - C_{10}^{\text{NP}} = \frac{\sqrt{2}}{4 G_F V_{tb} V_{ts}^*} \frac{y_s y_b^* |y_\mu|^2}{64 \pi \alpha m_\chi^2} (F(x_{\phi_q^{2/3}}, x_S) + F(x_{\phi_q^{2/3}}, x_A)) ,
	\label{eq:c9np}
\end{equation}
with 
\begin{equation}
	F(x, y) = \frac{1}{(1-x)(1-y)} + \frac{x^2 \ln x}{(1-x)^2(x-y)} + \frac{y^2 \ln y}{(1-y)^2(y-x)} \, \, .
\end{equation}
Using the most recent experimental results, the best fitted values of the Wilson coefficients are $C_9^{\text{NP}} = - C_{10}^{\text{NP}} = [-0.59, -0.30]$~\cite{Alguero:2021anc}, at a confidence level of 2$\sigma$. Only points in the parameter space that generate $C_9^{\text{NP}}$  within the 2$\sigma$ range around its central value are considered in our analysis. 

The final constraint that was considered is related to the $B_s-\overline{B}_s$ mixing. Once again, we introduce an effective hamiltonian to explain the $b \to s$ transition involved in this process,
\begin{equation}
	\mathcal{H}_{\text{eff}}^{B\overline{B}} = C_{B\overline{B}} \, (\overline{s}_\alpha \gamma^\mu P_L b_\alpha)(\overline{s}_\beta \gamma_\mu P_L b_\beta).
\end{equation}
The Wilson coefficient $C_{B\overline{B}}$ is given in our model by~\cite{2017}
\begin{equation}
	C_{B\overline{B}} = \frac{(y_s y_b^*)^2}{128 \pi^2 m_\chi^2} F(x_{\phi_q^{2/3}}, x_{\phi_q^{2/3}}), 
\end{equation}
with
\begin{equation}
	F(x, x) = \frac{1 - x^2 + 2x \ln x}{(1-x)^3} \, \, .
\end{equation}
The experimental limit is set by the mass difference, $\Delta M_s$, between $B_s$ and $\overline{B}_s$. Computing this difference in the SM, $\Delta M_s^{\text{SM}}$, and comparing it with its experimental counterpart, $\Delta M_s^{\text{exp}}$, we can define the quantity~\cite{arnan2021generic}
\begin{equation}
	R_{\Delta M_s} = \frac{\Delta M_s^{\text{exp}}}{\Delta M_s^{\text{SM}}} - 1 = -0.09 \pm 0.08 .
\end{equation}
This last expression can be rewritten as a function of the Wilson coefficients~\cite{arnan2021generic, GABBIANI1996321},  
\begin{equation}
	R_{\Delta M_s} = \lvert 1 + \frac{0.8 C_{B\overline{B}}(\mu_H)}{C_{B\overline{B}}^{\text{SM}}(\mu_b)} - 1 \rvert ,
\end{equation}
where $C_{B\overline{B}}(\mu_H)$ is the Wilson coefficient in our model defined at a high-energy scale of $\mu_H = 1$~TeV, and $C_{B\overline{B}}^{\text{SM}´}(\mu_b) \approx 7.2 \times 10^{-11}$~GeV$^{-2}$ the corresponding SM value at the scale $\mu_b$~\cite{PhysRevD.93.113016}. We will require $R_{\Delta M_s}$ to lie in its 2$\sigma$ range, thus constraining $C_{B\overline{B}}$.

\section{Dark Matter constraints}
\hspace{\parindent} 
\label{sec:4}

In this section, we discuss the constraints from DM physics. We have taken into account in our study 
the DM relic density observations, the constraints from DM direct detection and the collider searches.
As previously stated there are two scalars that could be the DM candidates. We have chosen the particle $S$ but we have checked that choosing $A$ would lead to exactly the same results. This is because particles $S$ and $A$ have exactly the same quantum
numbers except for their CP parities. Since these particles are in the dark sector ($Z_2$-odd) their CP is not determined and we can only say that they have opposite CP parities. This has no bearing in the interactions with the SM particles.
Being a DM candidate $S$  should reproduce the observed DM relic abundance, whose value is provided by the Planck Collaboration with $\Omega_{DM} h^2 = 0.120 \pm 0.001$~\cite{planck}. We assume that the ordinary freeze-out mechanism is responsible for the 
generation of the DM relic density. Thus, the number density of $S$, $n_S$, can be determined by solving the following Boltzmann equation
\begin{equation}
	\frac{dn_S}{dt} + 3 H n_S = - \langle \sigma v \rangle (n_S^2 - n_S^{\text{eq} \, 2}), 
	\label{eq:1}
\end{equation}
where $n^{\text{eq}}_S$ corresponds to the value of $n_S$ at equilibrium, $H$ is the Hubble parameter and $\langle \sigma v \rangle$ is the thermal average of the DM annihilation cross section times its relative velocity $v$. The Boltzmann Eq.~(\ref{eq:1}) is numerically solved using \texttt{MICROMEGAS}~\cite{2021} which takes all possible annihilation and coannihilation channels into account. The freeze-in mechanism, a well known alternative to explain the observed DM abundance, cannot be used in our model as it requires extremely weak couplings between the DM particle and the visible sector, of the order $\mathcal{O}(10^{-10} - 10^{-12})$~\cite{2018}. Consider for instance the contributing process for freeze-in $SS \to \mu^+ \mu^-$ with a cross section proportional to $|y_\mu|^2$. A very small $|y_\mu|$ would be allowed by the DM physics but it would not solve the muon $g-2$ discrepancy. 
 
\begin{figure}[!h]
	\begin{center}
		\begin{tikzpicture}
			\begin{feynman}
				\vertex (a) {\(S\)};
				\vertex [right=of a] (b);
				\vertex [right=of b] (c) {\(\mu^\pm\)};
				\vertex [below=0.75cm of b] (d);
				\vertex [right=0.05cm of d] {\(\chi^-\)};
				\vertex [below=0.75cm of d] (e);
				\vertex [left=of e] (f) {\(A, \phi_l^\pm, \chi^\pm\)};
				\vertex [right=of e] (g) {\(\mu^\mp, \nu_\mu, \gamma/Z\)};
				\diagram* {
					(a) -- [scalar] (b),
					(b) -- [] (c),
					(b) -- [] (e),
					(e) -- [] (f),
					(e) -- [] (g),
				};
			\end{feynman}
		\end{tikzpicture}
		\hspace{1cm}
		\begin{tikzpicture}
			\begin{feynman}
				\vertex (a) {\(S\)};
				\vertex [right=of a] (b);
				\vertex [right=of b] (c) {\(W^\pm\)};
				\vertex [below=0.75cm of b] (d);
				\vertex [right=0.05cm of d] {\(\phi_l^+\)};
				\vertex [below=0.75cm of d] (e);
				\vertex [left=of e] (f) {\(A, \phi_l^\pm, \chi^\pm\)};
				\vertex [right=of e] (g) {\(W^\mp, \gamma/Z/h, \nu_\mu\)};
				\diagram* {
					(a) -- [scalar] (b),
					(b) -- [boson] (c),
					(b) -- [scalar] (e),
					(e) -- [] (f),
					(e) -- [] (g),
				};
			\end{feynman}
		\end{tikzpicture}
		\\
		\vspace{0.5cm}
		\begin{tikzpicture}
			\begin{feynman}
				\vertex (a) {\(S\)};
				\vertex [right=of a] (b);
				\vertex [right=of b] (c) {\(Z\)};
				\vertex [below=0.75cm of b] (d);
				\vertex [right=0.05cm of d] {\(A\)};
				\vertex [below=0.75cm of d] (e);
				\vertex [left=of e] (f) {\(A, \phi_l^\pm, \chi^\pm\)};
				\vertex [right=of e] (g) {\(h, W^\pm, \mu^\pm\)};
				\diagram* {
					(a) -- [scalar] (b),
					(b) -- [boson] (c),
					(b) -- [scalar] (e),
					(e) -- [] (f),
					(e) -- [] (g),
				};
			\end{feynman}
		\end{tikzpicture}
		\hspace{1cm}
		\vspace{0.5cm}
		\begin{tikzpicture}
			\begin{feynman}
				\vertex (a) {\(S\)};
				\vertex [below right=of a] (b);
				\vertex [below left=of b] (c) {\(A, \phi_l^\pm, \chi^\pm\)};
				\vertex [right=1.5cm of b] (d);
				\vertex [above=0.05cm of d] {\(Z, W^\pm, \mu^\pm\)};
				\vertex [right=1.5cm of d] (e);
				\vertex [above right=of e] (f) {\(\)};
				\vertex [below right=of e] (g) {\(\)};
				\diagram* {
					(a) -- [scalar] (b),
					(b) -- [] (c),
					(b) -- [] (e),
					(e) -- [] (f),
					(e) -- [] (g),
				};
			\end{feynman}
		\end{tikzpicture}
		\\
		\begin{tikzpicture}
			\begin{feynman}
				\vertex (a) {\(S\)};
				\vertex [below right=of a] (b);
				\vertex [below left=of b] (c) {\(\phi_l^\pm\)};
				\vertex [above right=of b] (d) {\(\gamma/Z\)};
				\vertex [below right=of b] (e) {\(W^\pm\)};
				\diagram* {
					(a) -- [scalar] (b),
					(b) -- [scalar] (c),
					(b) -- [boson] (d),
					(b) -- [boson] (e),
				};
			\end{feynman}
		\end{tikzpicture}
		\caption{Feynman diagrams for DM coannihilation in Model 3. The solid lines without arrows represent scalars, gauge bosons or fermions. Different initial and final states are separated by a comma. The notation $x/y$ means that both particles can exist in the final state, for a given initial state.
		}
		\label{fig:dmcoani}
	\end{center}
\end{figure}
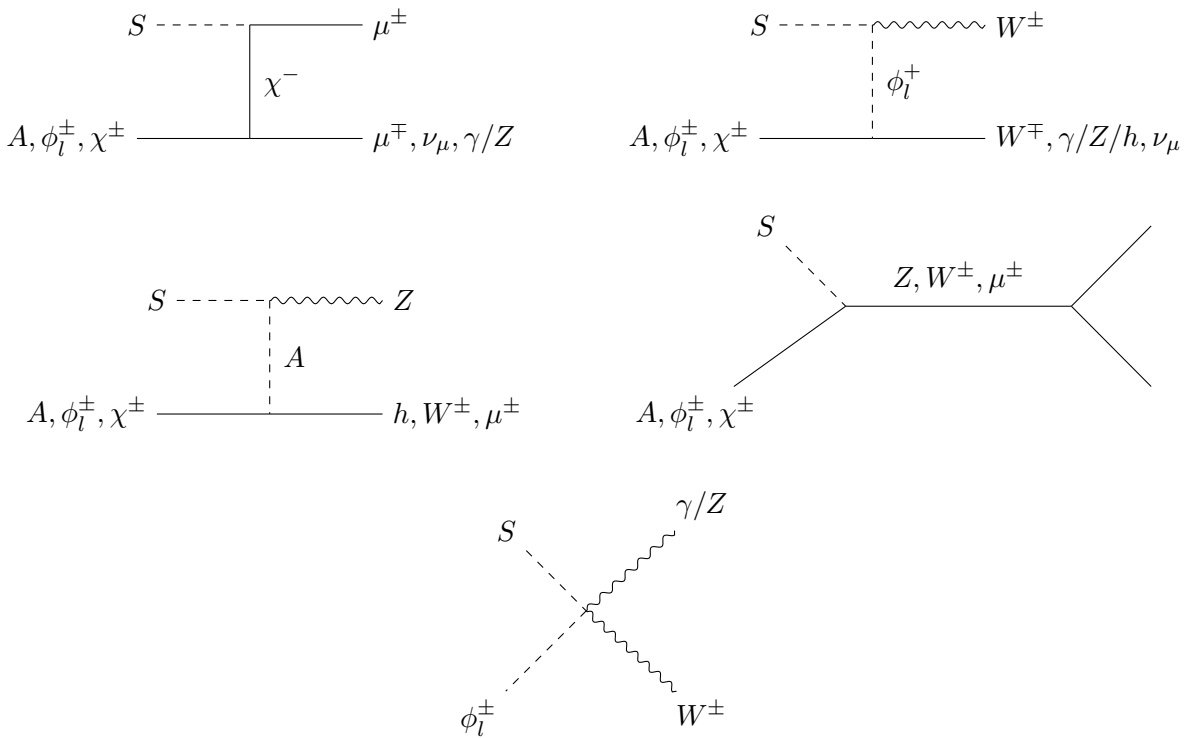

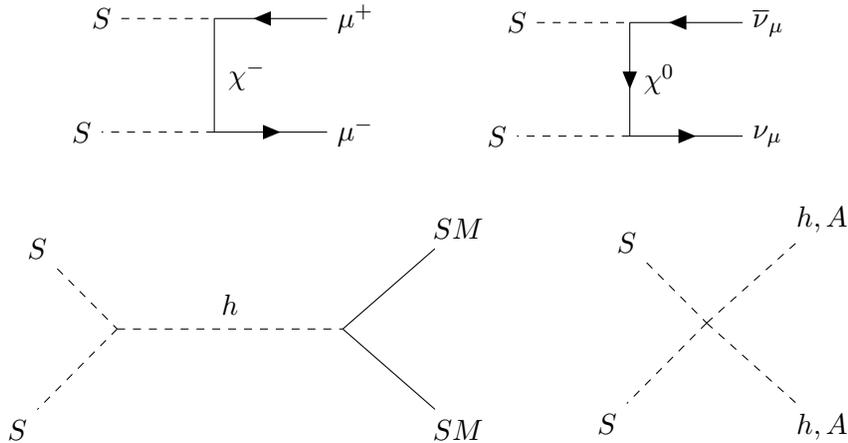
\begin{figure}[!h]
	\begin{center}
		\begin{tikzpicture}
			\begin{feynman}
				\vertex (a) {\(S\)};
				\vertex [right=of a] (b);
				\vertex [right=of b] (c) {\(\mu^+\)};
				\vertex [below=0.75cm of b] (d);
				\vertex [right=0.05cm of d] {\(\chi^-\)};
				\vertex [below=0.75cm of d] (e);
				\vertex [left=of e] (f) {\(S\)};
				\vertex [right=of e] (g) {\(\mu^-\)};
				\diagram* {
					(a) -- [scalar] (b),
					(b) -- [anti fermion] (c),
					(b) -- [] (e),
					(e) -- [scalar] (f),
					(e) -- [fermion] (g),
				};
			\end{feynman}
		\end{tikzpicture}
		\hspace{1cm}
		\begin{tikzpicture}
			\begin{feynman}
				\vertex (a) {\(S\)};
				\vertex [right=of a] (b);
				\vertex [right=of b] (c) {\(\overline{\nu}_\mu\)};
				\vertex [below=0.75cm of b] (d);
				\vertex [right=0.05cm of d] {\(\chi^0\)};
				\vertex [below=0.75cm of d] (e);
				\vertex [left=of e] (f) {\(S\)};
				\vertex [right=of e] (g) {\(\nu_\mu\)};
				\diagram* {
					(a) -- [scalar] (b),
					(b) -- [anti fermion] (c),
					(b) -- [fermion] (e),
					(e) -- [scalar] (f),
					(e) -- [fermion] (g),
				};
			\end{feynman}
		\end{tikzpicture}
		\\
		\vspace{0.5cm}
		\begin{tikzpicture}
			\begin{feynman}
				\vertex (a) {\(S\)};
				\vertex [below right=of a] (b);
				\vertex [below left=of b] (c) {\(S\)};
				\vertex [right=1.5cm of b] (d);
				\vertex [above=0.05cm of d] {\(h\)};
				\vertex [right=1.5cm of d] (e);
				\vertex [above right=of e] (f) {\(SM\)};
				\vertex [below right=of e] (g) {\(SM\)};
				\diagram* {
					(a) -- [scalar] (b),
					(b) -- [scalar] (c),
					(b) -- [scalar] (e),
					(e) -- [] (f),
					(e) -- [] (g),
				};
			\end{feynman}
		\end{tikzpicture}
		\hspace{1cm}
		\begin{tikzpicture}
			\begin{feynman}
				\vertex (a) {\(S\)};
				\vertex [below right=of a] (b);
				\vertex [below left=of b] (c) {\(S\)};
				\vertex [above right=of b] (d) {\(h, A\)};
				\vertex [below right=of b] (e) {\(h, A\)};
				\diagram* {
					(a) -- [scalar] (b),
					(b) -- [scalar] (c),
					(b) -- [scalar] (d),
					(b) -- [scalar] (e),
				};
			\end{feynman}
		\end{tikzpicture}
		\caption{Feynman diagrams for DM annihilation in Model 5. The solid lines without arrows represent scalars, gauge bosons or fermions. Different initial and final states are separated by a comma. The notation $x/y$ means that both particles can exist in the final state, for a given initial state. "SM" represents all massive SM particles. In Model 5, $\chi$ is a fermion doublet with $\chi = (\chi^0, \chi^-).$}
		\label{fig:dmani5}
	\end{center}
\end{figure}

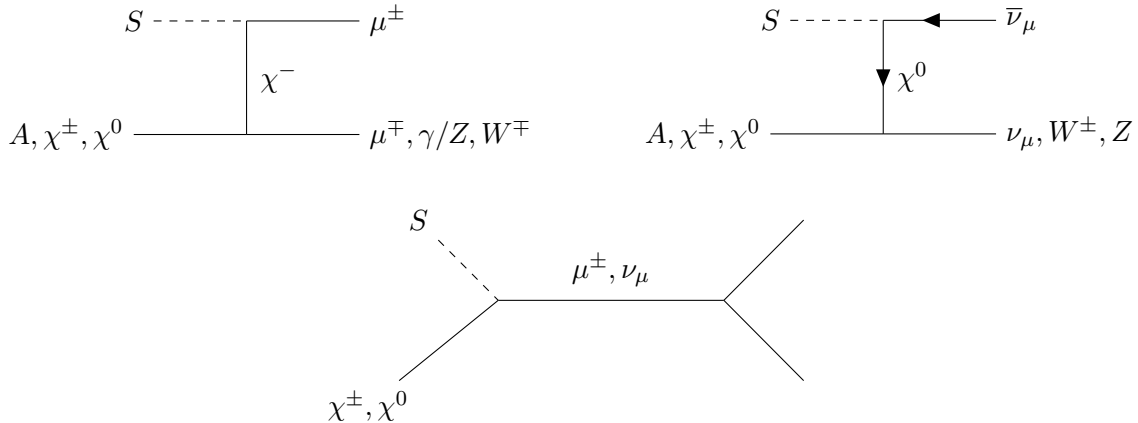
\begin{figure}[!h]
	\begin{center}
		\begin{tikzpicture}
			\begin{feynman}
				\vertex (a) {\(S\)};
				\vertex [right=of a] (b);
				\vertex [right=of b] (c) {\(\mu^\pm\)};
				\vertex [below=0.75cm of b] (d);
				\vertex [right=0.05cm of d] {\(\chi^-\)};
				\vertex [below=0.75cm of d] (e);
				\vertex [left=of e] (f) {\(A, \chi^\pm, \chi^0\)};
				\vertex [right=of e] (g) {\(\mu^\mp, \gamma/Z, W^\mp\)};
				\diagram* {
					(a) -- [scalar] (b),
					(b) -- [] (c),
					(b) -- [] (e),
					(e) -- [] (f),
					(e) -- [] (g),
				};
			\end{feynman}
		\end{tikzpicture}
		\hspace{1cm}
		\begin{tikzpicture}
			\begin{feynman}
				\vertex (a) {\(S\)};
				\vertex [right=of a] (b);
				\vertex [right=of b] (c) {\(\overline{\nu}_\mu\)};
				\vertex [below=0.75cm of b] (d);
				\vertex [right=0.05cm of d] {\(\chi^0\)};
				\vertex [below=0.75cm of d] (e);
				\vertex [left=of e] (f) {\(A, \chi^\pm, \chi^0\)};
				\vertex [right=of e] (g) {\(\nu_\mu, W^\pm, Z\)};
				\diagram* {
					(a) -- [scalar] (b),
					(b) -- [anti fermion] (c),
					(b) -- [fermion] (e),
					(e) -- [] (f),
					(e) -- [] (g),
				};
			\end{feynman}
		\end{tikzpicture}
		\\
		\vspace{0.5cm}
		\begin{tikzpicture}
			\begin{feynman}
				\vertex (a) {\(S\)};
				\vertex [below right=of a] (b);
				\vertex [below left=of b] (c) {\(\chi^\pm, \chi^0\)};
				\vertex [right=1.5cm of b] (d);
				\vertex [above=0.05cm of d] {\(\mu^\pm, \nu_\mu\)};
				\vertex [right=1.5cm of d] (e);
				\vertex [above right=of e] (f) {\(\)};
				\vertex [below right=of e] (g) {\(\)};
				\diagram* {
					(a) -- [scalar] (b),
					(b) -- [] (c),
					(b) -- [] (e),
					(e) -- [] (f),
					(e) -- [] (g),
				};
			\end{feynman}
		\end{tikzpicture}
		\caption{Feynman diagrams for DM coannihilation in Model 5. The solid lines without arrows represent scalars, gauge bosons or fermions. Different initial and final states are separated by a comma. The notation $x/y$ means that both particles can exist in the final state, for a given initial state. In Model 5, $\chi$ is a fermion doublet with $\chi = (\chi^0, \chi^-).$}
		\label{fig:dmcoani5}
	\end{center}
\end{figure}

 In Figures~\ref{fig:dmani} and \ref{fig:dmcoani}, we show the relevant Feynman diagrams that contribute to the main processes of DM annihilation and coannihilation, respectively, in Model 3. The corresponding set of diagrams for Model 5 is shown in Figures~\ref{fig:dmani5} and \ref{fig:dmcoani5}. A key aspect shown in these diagrams is that, since the scalar fields are doublets in Model 3, they can couple to gauge bosons, unlike what happens in Model 5 where the scalar fields are singlets. This will drastically change the distribution of the DM relic abundance, as will be shown further ahead. 

\begin{figure}[!h]
	\begin{center}
		\begin{tikzpicture}
			\begin{feynman}
				\vertex (a) {\(S\)};
				\vertex [right=of a] (b);
				\vertex [right=of b] (c) {\(\mu^+\)};
				\vertex [below=0.75cm of b] (d);
				\vertex [right=0.05cm of d] {\(\chi^-\)};
				\vertex [below=0.75cm of d] (e);
				\vertex [left=of e] (f) {\(S\)};
				\vertex [right=of e] (g) {\(\mu^-\)};
				\diagram* {
					(a) -- [scalar] (b),
					(b) -- [anti fermion] (c),
					(b) -- [] (e),
					(e) -- [scalar] (f),
					(e) -- [fermion] (g),
				};
			\end{feynman}
		\end{tikzpicture}
		\hspace{1cm}
		\begin{tikzpicture}
			\begin{feynman}
				\vertex (a) {\(S\)};
				\vertex [right=of a] (b);
				\vertex [right=of b] (c) {\(W^+\)};
				\vertex [below=0.75cm of b] (d);
				\vertex [right=0.05cm of d] {\(\phi_l^+\)};
				\vertex [below=0.75cm of d] (e);
				\vertex [left=of e] (f) {\(S\)};
				\vertex [right=of e] (g) {\(W^-\)};
				\diagram* {
					(a) -- [scalar] (b),
					(b) -- [boson] (c),
					(b) -- [scalar] (e),
					(e) -- [scalar] (f),
					(e) -- [boson] (g),
				};
			\end{feynman}
		\end{tikzpicture}
		\hspace{1cm}
		\begin{tikzpicture}
			\begin{feynman}
				\vertex (a) {\(S\)};
				\vertex [right=of a] (b);
				\vertex [right=of b] (c) {\(Z\)};
				\vertex [below=0.75cm of b] (d);
				\vertex [right=0.05cm of d] {\(A\)};
				\vertex [below=0.75cm of d] (e);
				\vertex [left=of e] (f) {\(S\)};
				\vertex [right=of e] (g) {\(Z\)};
				\diagram* {
					(a) -- [scalar] (b),
					(b) -- [boson] (c),
					(b) -- [scalar] (e),
					(e) -- [scalar] (f),
					(e) -- [boson] (g),
				};
			\end{feynman}
		\end{tikzpicture}
		\\
		\vspace{0.5cm}
		\begin{tikzpicture}
			\begin{feynman}
				\vertex (a) {\(S\)};
				\vertex [below right=of a] (b);
				\vertex [below left=of b] (c) {\(S\)};
				\vertex [right=1.5cm of b] (d);
				\vertex [above=0.05cm of d] {\(h\)};
				\vertex [right=1.5cm of d] (e);
				\vertex [above right=of e] (f) {\(SM\)};
				\vertex [below right=of e] (g) {\(SM\)};
				\diagram* {
					(a) -- [scalar] (b),
					(b) -- [scalar] (c),
					(b) -- [scalar] (e),
					(e) -- [] (f),
					(e) -- [] (g),
				};
			\end{feynman}
		\end{tikzpicture}
		\hspace{1cm}
		\begin{tikzpicture}
			\begin{feynman}
				\vertex (a) {\(S\)};
				\vertex [below right=of a] (b);
				\vertex [below left=of b] (c) {\(S\)};
				\vertex [above right=of b] (d) {\(h, A, \phi_l^+, Z, W^+\)};
				\vertex [below right=of b] (e) {\(h, A, \phi_l^-, Z, W^-\)};
				\diagram* {
					(a) -- [scalar] (b),
					(b) -- [scalar] (c),
					(b) -- [] (d),
					(b) -- [] (e),
				};
			\end{feynman}
		\end{tikzpicture}
		\caption{Feynman diagrams for DM annihilation in Model 3. The solid lines without arrows represent scalars, gauge bosons or fermions. Different initial and final states are separated by a comma. The notation $x/y$ means that both particles can exist in the final state, for a given initial state. "SM" represents all massive SM particles.
		}
		\label{fig:dmani}
	\end{center}
\end{figure}
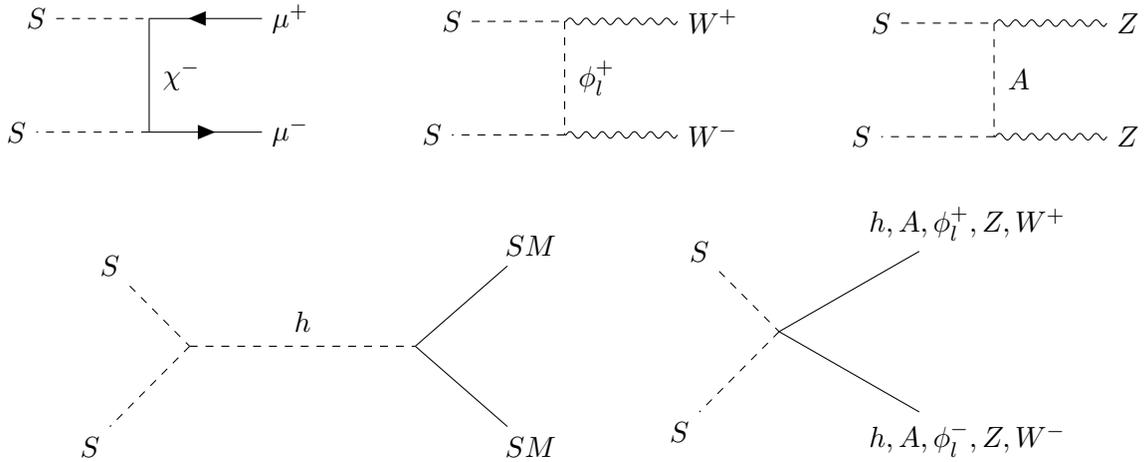

Besides the DM relic abundance, DM direct detection (DD) may also place severe constraints on the parameter space of Model 3. Currently, the best experimental upper bounds on the DM direct detection cross section for a mass above 6~GeV are provided by the PandaX-4T~\cite{PandaX-4T:2021bab} and by the XENON1T~\cite{PhysRevLett.121.111302} experiments. Very recently the LuxZeplin (LZ) experiment has also released their bounds on the spin independent cross section, the best so far~\cite{LZ:2021xov, LZnew}. We will show the three limits in our plots. This will allow to understand the effect of future DD bounds. 

In Model 3, the tree-level t-channel diagram corresponding to the process $S N \to S N$ (where $N$ is a nucleon), mediated by the SM Higgs boson, represents the main contribution to the DM-nucleon scattering cross section, given by 
\begin{equation}
	\sigma (S N \to S N) = \frac{(\lambda_{12} + \lambda_{12}' + 2 \lambda_{5})^2}{4 \pi} \frac{f_N^2 m_N^2 \mu_{SN}^2}{m_S^2 m_h^4} \, ,
	\label{eq:2}
\end{equation}
where $m_N$ is the nucleon mass, $\mu_{SN}$ is the reduced mass of the DM-nucleon pair, and $f_N \approx 0.3$ is the effective Higgs-nucleon coupling~\cite{PhysRevD.88.055025, PhysRevD.85.051503, REN20187}. Furthermore, we also consider the limits coming from collider searches at the LHC. In particular, we take the constraint from the SM-like Higgs boson invisible decay into an $S$ pair. The invisible decay width in our model, valid for $m_S < m_h/2$, is given by
\begin{equation}
	\Gamma (h \to S S) = \frac{(\lambda_{12} + \lambda_{12}' + 2 \lambda_{5})^2 v^2}{32 \pi m_h} \sqrt{ 1- \frac{4 m_S^2}{m_h^2}} .
	\label{eq:3}
\end{equation}
The upper limit for the Higgs to invisible branching ratio is provided by the LHC, with $\mathcal{B} (h \to S S) \leq 0.11$~\cite{Zyla:2020zbs}. 
With the present constraints,
DM direct detection experiments give rise to much tighter bounds than the Higgs invisible width. We will come back to this point later.

\section{Results}
\label{sec:5}
\subsection{Initial scan setup}
\hspace{\parindent} 
\label{section51}

In this section we discuss the results obtained for Model 3 when taking into account the previously mentioned flavor and DM constraints, by performing multiparameter scans to identify the allowed parameter space. 

The relevant input parameters for Model 3 are:
\begin{equation}
	y_b, \, y_s, \, y_\mu, \, m_\chi, m_{\phi_q^{5/3}}, \, m_{\phi_q^{2/3}}, \, m_S, \, m_A, \, m_{\phi_{l}}, \, \lambda_{12} \, (\text{or} \, \lambda_{hS}), \, \lambda_2,
	\label{eq:inputMod3}
\end{equation}
where $\lambda_{hS} \equiv \lambda_{12} + \lambda_{12}' + 2 \lambda_{5}$ is the Higgs portal coupling. 
In principle, the free input quartic parameters from the Higgs potential in Eq.~(\ref{eq:pot}), $\lambda_{23}$, $\lambda_{23}'$ and $y_{23}$ could also be relevant for the discussion, since they contribute to the DM abundance via coannihilation channels involving the coloured scalar fields. However, since the mass difference between $S$ and $\phi_{q}^{\pm 5/3}$ (or $\phi_{q}^{\pm 2/3}$) is very large, the contribution of these channels to the total relic density is greatly suppressed. 
We set $\lambda_{23} = \lambda_{23}' = y_{23} = 10^{-3}$, but we could have chosen much larger values. The parameters $\lambda_3$, $\lambda_{13}$ and $y_{13}$ are also irrelevant for the discussion that follows, since they do not contribute to any of the flavor and DM physics that we wish to explain.

The results we will show next consist of two different scans. In scan I (Figures~\ref{fig:DMrelic} and \ref{fig:portal}), we tried to get a feel for the allowed parameter space of Model 3, by naively varying its relevant input parameters in a very similar fashion as what was done for Model 5. In the second scan, scan II, we fine-tuned the parameters using what we learned from scan I, in order to make sure that we had points satisfying all the previously mentioned constraints. The results from scan II are our final and main results. In our figures, all points of the parameter space explain the $B$ meson data within its 2$\sigma$ confidence intervals. The cyan points are excluded when taking into account the observed DM relic abundance in the 2$\sigma$ range. The blue points do not satisfy the constraints from DM direct detection (from XENON1T) and collider searches. The green points are not allowed by the muon $(g-2)$ data within its 3$\sigma$ range, and finally, the red points represent the parameter space which can explain all the flavor and DM constraints simultaneously.

Before we move on to the actual results, we need to state and explain first the values considered for the parameters in Eq.~(\ref{eq:inputMod3}), for both scans. Following the arguments in~\cite{PhysRevD.102.075009}, since only the combination $y_s y_b^*$ appears in the equations related to the $B$ meson decays and $B_s-\overline{B}_s$ mixing, which should be negative to solve the deficit in the measurements of $R(K^*)$, both $y_b$ and $y_s$ are real and proportional to each other, with $y_s = - y_b/4$. We set $|y_b| \leq 1$ (both scans), $0 \leq y_\mu \leq 4 \pi$ (scan I) and $1 \leq y_\mu \leq 4 \pi$ (scan II). We stress that the flavor physics is the same in Models 3 and 5, thus it is reasonable to vary in a similar way these parameters.
The condition $1 \leq y_\mu$ in scan II is just for optimization purposes.

The masses of the coloured scalars are fixed and equal to 1.5~TeV like in Model 5. The remaining particles in the dark sector are forced to be heavier than the DM particle $S$ by at least 10~GeV (and at most 1~TeV) in both scans, with 5~GeV $\leq m_S \leq 1$~TeV, the standard WIMP range (scan I), and 5~GeV $\leq m_S \leq 100$~GeV (scan II). The lower upper limit for the DM mass in scan II is to optimize the scan, since for reasons we will explain ahead $m_S$ is constrained to be below roughly 80~GeV to satisfy the DM constraints. For the masses of the other colourless scalars, we have, in scan I, 15~GeV $\leq m_A \leq 2$~TeV and 15~GeV $\leq m_{\phi_{l}} \leq 2$~TeV. In scan II, we considered additional constraints coming from precision data and LEP experiments from the measurements of the $W$ and $Z$ boson widths. In order for the decay channels $W^\pm \to S \phi_{l}^\pm, A \phi_{l}^\pm$ and $Z \to S A, \phi_{l}^+ \phi_{l}^-$ to be kinematically forbidden, the following lower limits must be obeyed
\begin{equation}
	m_S + m_{\phi_{l}} > m_W \, , \quad m_A + m_{\phi_{l}} > m_W \, , \quad m_S + m_A > m_Z \, , \quad 2 m_{\phi_{l}} > m_Z . 
	\label{eq:LEPcons}
\end{equation} 
\begin{table}[h!]
	\renewcommand{\arraystretch}{2.}
	\begin{center}
		\begin{tabular}{ |c|c|c|c|c| } 
			\hline
			$y_b$ & $y_s$ & $y_\mu$ & $m_\chi$ (GeV) & $m_{\phi_q^{5/3}}$ and $m_{\phi_q^{2/3}}$ (GeV) \\ \hline
			[-1, 1] & $-y_b/4$ & [0, $4\pi$] & [101.2, 2000] & 1500 \\
			\hline \hline
			$m_S$ (GeV) & $m_A$ (GeV) & $m_{\phi_{l}}$ (GeV) & $\lambda_{12}$ & $\lambda_2$ \\ \hline
			[5, 1000] & [15, 2000] & [15, 2000] & [$10^{-5}$, 0.5] & $10^{-5}$ \\
			\hline
		\end{tabular}
	\end{center}
	\caption{Values assigned to the relevant Model 3 input parameters in scan I. We further impose $|\lambda_{hS}| \leq 1$, which is achieved by rejecting points where $\lambda_5 < -0.2$ and $|\lambda_{12}'| \geq 0.1$.}
	\label{table:scanI}
\end{table}
Additionally, $e^+ e^- \to \phi_{l}^+ \phi_{l}^-$ at LEP sets the limit $ m_{\phi_{l}} > 70$~GeV~\cite{Pierce_2007}, and the regions defined by the conditions $m_S < 80$~GeV, $m_A < 100$~GeV and $m_A - m_S > 8$~GeV are also excluded by LEP since they would result in a visible di-jet or di-lepton signal~\cite{Lundstr_m_2009}. Thus, we have 100~GeV $\leq m_A \leq 1.1$~TeV and 70~GeV $\leq m_{\phi_{l}} \leq 1.1$~TeV in scan II. 
\begin{table}[h!]
	\renewcommand{\arraystretch}{2.}
	\begin{center}
		\begin{tabular}{ |c|c|c|c|c| } 
			\hline
			$y_b$ & $y_s$ & $y_\mu$ & $m_\chi$ (GeV) & $m_{\phi_q^{5/3}}$ and $m_{\phi_q^{2/3}}$ (GeV) \\ \hline
			[-1, 1] & $-y_b/4$ & [1, $4\pi$] & [101.2, 1100] & 1500 \\
			\hline \hline
			$m_S$ (GeV) & $m_A$ (GeV) & $m_{\phi_{l}}$ (GeV) & $\lambda_{12}$ & $\lambda_2$ \\ \hline
			[5, 100] & [100, 1100] & [70, 1100] & $\leq 4 \pi$ & $10^{-5}$ \\
			\hline
		\end{tabular}
	\end{center}
	\caption{Values assigned to the relevant Model 3 input parameters in scan II. We further impose $10^{-7} \leq |\lambda_{hS}| \leq 10^{-2}$, and $|\lambda_{5}|$ and $|\lambda_{12}'| \leq 4\pi$.}
	\label{table:scanII}
\end{table}
It turned out that the constraints from Eq.~(\ref{eq:LEPcons}) do not need to be imposed because they are automatically satisfied for the points that verify all constraints (red points). Also, we impose $m_A \geq 100$~GeV initially since we found from scan I that in the allowed parameter region we must have $m_S < 80$~GeV (and $m_A - m_S > 8$~GeV is satisfied by design). Furthermore, the masses of $S$, $A$ and $\phi_l$ must be such that the dimensionless couplings $\lambda_{12}'$ and $\lambda_{5}$ are smaller than their perturbative limit of $4\pi$ (both scans). For the vectorlike fermion, we set 101.2~GeV $\leq m_\chi \leq 2$~TeV (scan I) and 101.2~GeV $\leq m_\chi \leq 1.1$~TeV (scan II). The lower limit on $m_\chi$ comes from LEP searches for unstable heavy vectorlike charged leptons~\cite{ACHARD200175}. More recent constraints from the LHC exist for vectorlike leptons, but they do not apply to our model since those searches assume that the vectorlike leptons couple to tau leptons~\cite{Sirunyan_2019}, or very small amounts of missing transverse energy, $\cancel{\it{E}}_{T}$, in the final states~\cite{Bi_mann_2021}. Regarding the Higgs portal coupling, we impose $|\lambda_{hS}| \leq 1$ in scan I, like we did in Model 5, which is achieved by setting $10^{-5} \leq \lambda_{12} \leq 0.5$ and rejecting points where $\lambda_5 < -0.2$ and $|\lambda_{12}'| \geq 0.1$. In scan II, $10^{-7} \leq |\lambda_{hS}| \leq 10^{-2}$, and $\lambda_{12}$, $|\lambda_{5}|$ and $|\lambda_{12}'| \leq 4\pi$. 
Since the Higgs portal coupling depends on the masses of $S$ and $\phi_{l}$ in Model 3, unlike Model 5 where it is a completely free parameter, $\lambda_{12}$ needs to be fine-tuned for $\lambda_{hS}$ to be very small. This will be discussed further ahead. Finally we have $\lambda_2$, whose only contribution is to the DM relic abundance, through the channels $SS \to AA, \phi_{l}^+ \phi_{l}^-$. We set $\lambda_2 = 10^{-5}$ in both scans, to suppress its contribution to the relic density. We did not vary $\lambda_2$ in any of the scans, but we checked that we can have points satisfying the Planck observations for much larger values of $\lambda_2$. A summary of the values used for each relevant parameter in scans I and II is shown in Tables~\ref{table:scanI} and \ref{table:scanII}, respectively. 

\subsection{Model 3 vs Model 5}
\hspace{\parindent} 

\begin{figure}[h!]
	\begin{center}
		\hspace*{-5mm}
		\begin{tabular}{c c}
		\hspace*{10mm} \textbf{Model 3} & \hspace*{10mm} \textbf{Model 5} \\
		\includegraphics[height = 7.cm]{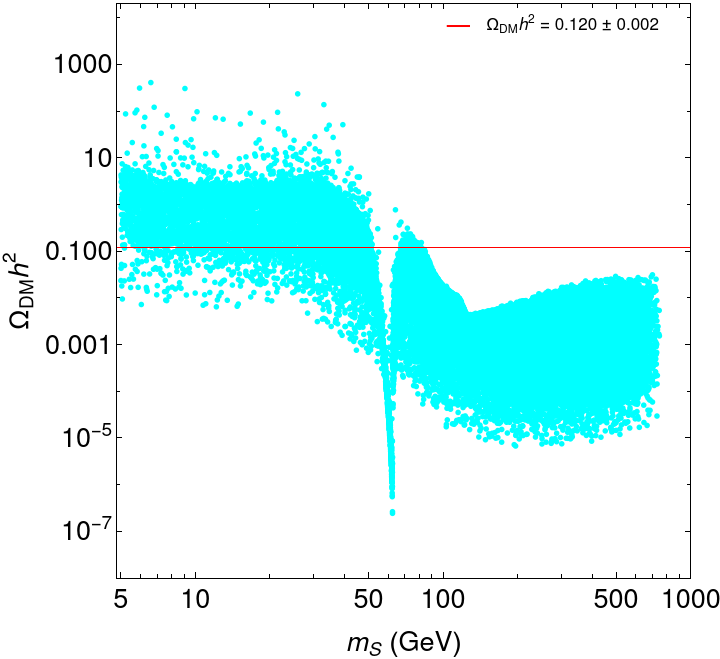} &
		\includegraphics[height = 7.cm]{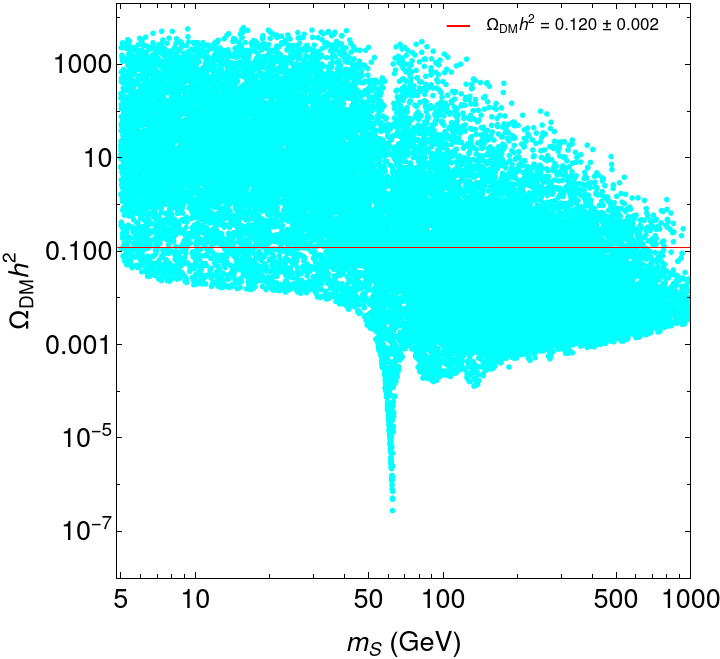}
		\end{tabular}
		\caption{DM relic density as a function of its mass for Model 3 (left) and Model 5 (right). The solid red line represents the 2$\sigma$ region for the observed DM relic abundance. The cyan points satisfy the $B$ meson data within its 2$\sigma$ confidence intervals, but are excluded when considering the observed DM relic density within the 2$\sigma$ CL range. The plot on the left was obtained from scan I. On the right, the values used for every Model 5 parameter are the same as in~\cite{PhysRevD.102.075009}.}
		\label{fig:DMrelic}
	\end{center}
\end{figure} 

In both models, there are regions of the parameter space satisfying all the constraints that were considered. However, there are differences in the allowed parameter space. The main difference between Models 3 and 5 is related to the DM's relic density distribution as a function of its mass. That can be seen in Figure \ref{fig:DMrelic} (scan I), where for Model 3 (left), we must have $m_S < 80$~GeV, roughly, to be close to the Planck observations. For Model 5 (right), no such limit is observed. The scalar fields have different $SU(2)_L$ representations: they are doublets in Model 3, and singlets in Model 5. This difference allows the scalar fields in Model 3 to couple to gauge bosons, and in particular for the annihilation processes $SS \to W^+ W^-$ and $SS \to ZZ$ to exist, which does not happen in Model 5 (see Figures~\ref{fig:dmani} and \ref{fig:dmani5}). This results in a DM relic abundance for Model 3 always smaller than the one given by the Planck observations when $m_S \geq m_W$, similar to what we observe for the i2HDM~\cite{Belyaev_2018}.

\begin{figure}[h!]
	\begin{center}
		\hspace*{-5mm}
		\begin{tabular}{c c}
		\hspace*{10mm} \textbf{Model 3} & \hspace*{10mm} \textbf{Model 5} \\
		\includegraphics[height = 7.cm]{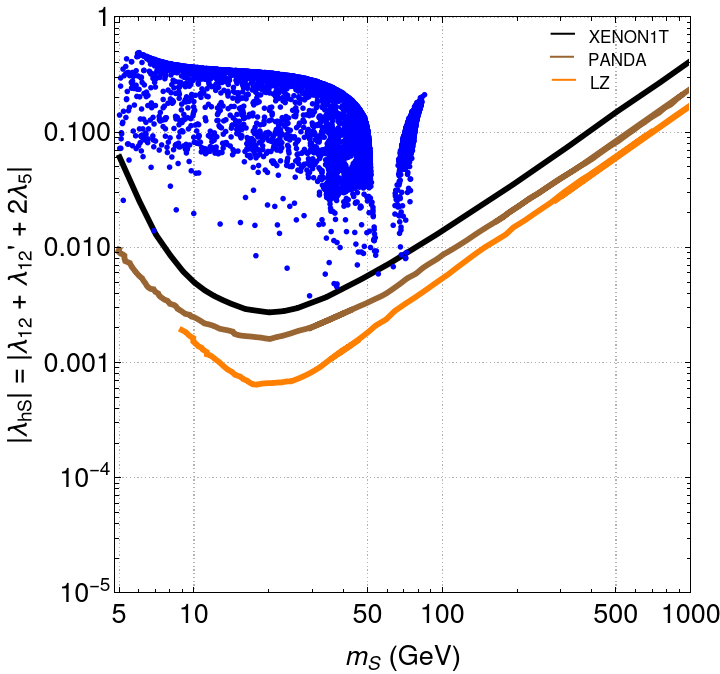} & \includegraphics[height = 7.cm]{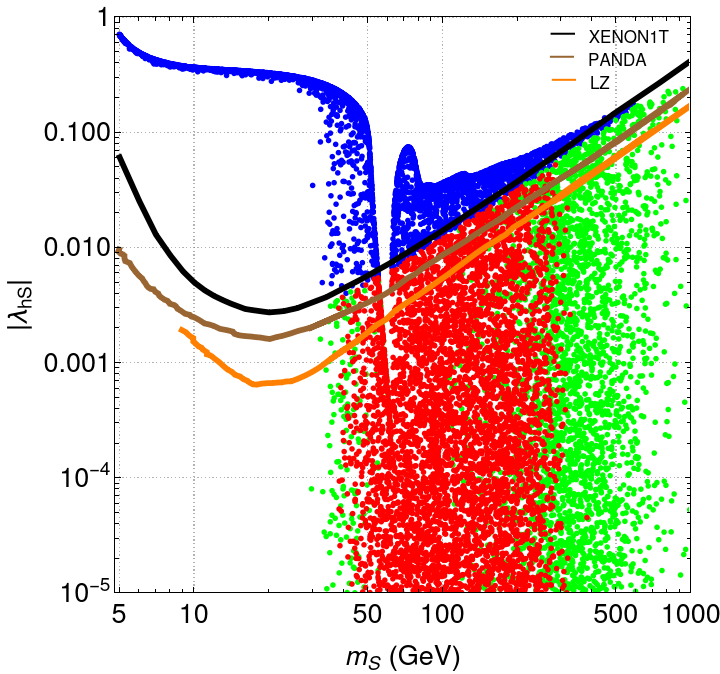}
		\end{tabular}
		\caption{Scan I - Higgs portal couplings $|\lambda_{hS}|$ as a function of the DM mass for Model 3 (left) and Model 5 (right). The solid black, brown and orange lines represent the experimental upper bounds provided by XENON1T, PandaX-4T and LZ, respectively. The blue points satisfy the $B$ meson data within its 2$\sigma$ confidence intervals, the observed DM relic density within its 2$\sigma$ CL range, but are excluded when considering DM direct detection and Higgs decays to invisible constraints. The green points satisfy all constraints except for the $(g - 2)$ of the muon, and the red points satisfy all constraints. The values used for the different Model 3 and 5 parameters are the same as in Figure~\ref{fig:DMrelic}.}
		\label{fig:portal}
	\end{center}
\end{figure}

Another distinction between Models 3 and 5 is associated to the Higgs portal coupling. In Model 3, that coupling is not a free input parameter like in Model 5, where it can be as small as needed. Since $\lambda_{hS} = \lambda_{12} + \lambda_{12}' + 2 \lambda_{5} = \lambda_{12} + 2 \, (m_S^2 - m_{\phi_{l}}^2)/v^2$, in order to have small values of $\lambda_{hS}$ in Model 3, the difference between $m_S$ and $m_{\phi_{l}}$ must also be small, or $\lambda_{12}$ must be close to $- 2 \, (m_S^2 - m_{\phi_{l}}^2)/v^2$. The reason for $\lambda_{hS}$ to be small (in the order of $10^{-2}$ or lower), is imposed by the experimental upper bound of the LZ, PandaX-4T and XENON1T experiments. 
%
This is shown in Figure~\ref{fig:portal} (scan I). As opposed to what we see on the plot on the right (Model 5), 
all points on the left plot (Model 3) are excluded due to the DM direct detection and Higgs invisible decays constraints. Given that $m_S$ varies between [5, 1000]~GeV, and the minimum mass difference between the DM candidate and any other new particle is 10~GeV, $|2 \, (m_S^2 - m_{\phi_{l}}^2)/v^2|$ can only be as small as $\approx 0.0066$. Thus, the necessary condition $\lambda_{hS} \leq 10^{-2}$ will be extremely unlikely to occur without forcing $\lambda_{12} \approx - 2 \, (m_S^2 - m_{\phi_{l}}^2)/v^2$ or $m_S - m_{\phi_{l}}$ to be smaller. From these two possibilities to keep $\lambda_{hS}$ small, we chose the former, since decreasing the mass difference between particles of the dark sector makes the coannihilation processes more efficient.

\begin{figure}[h!]
	\begin{center}
		\hspace{-5mm}
		\includegraphics[height = 6.cm]{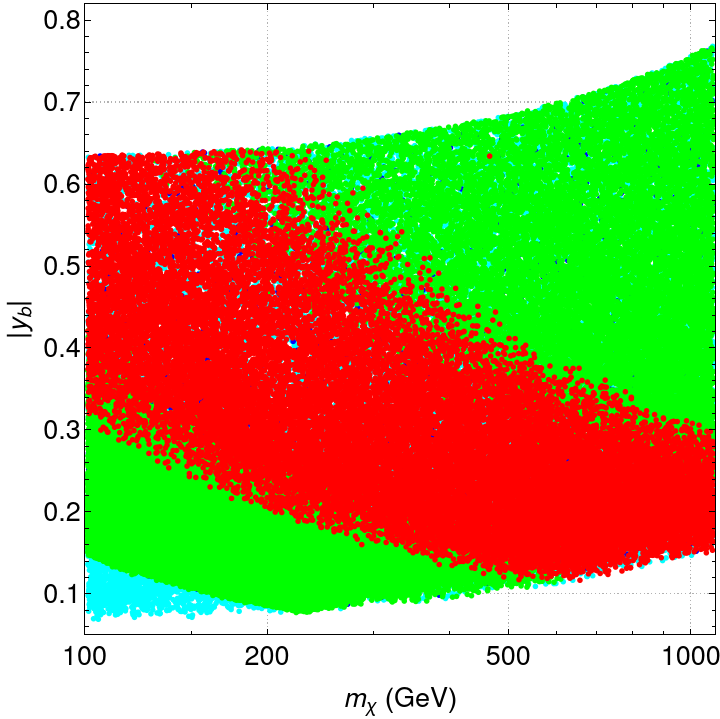}
		\hspace{2mm}
		\includegraphics[height = 6.cm]{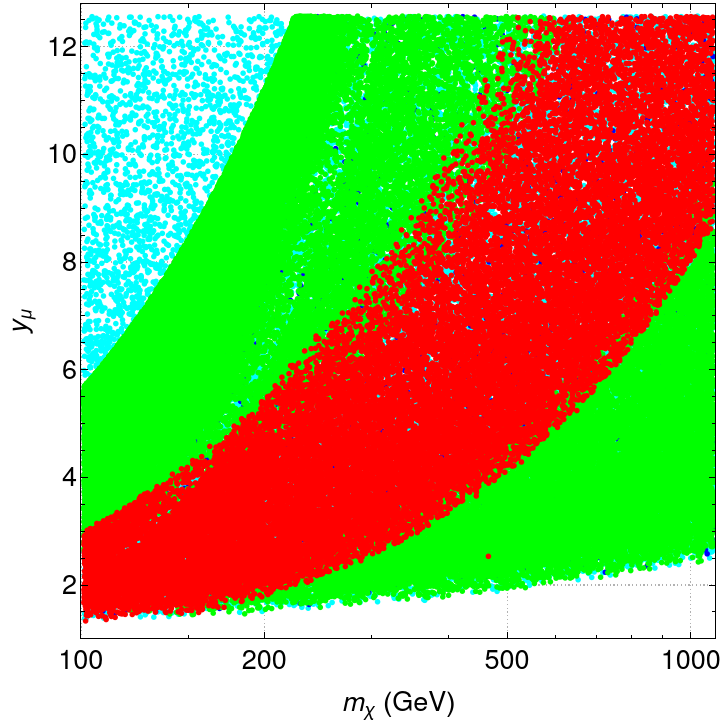}
		\\
		\vspace{2mm}
		\includegraphics[height = 6.cm]{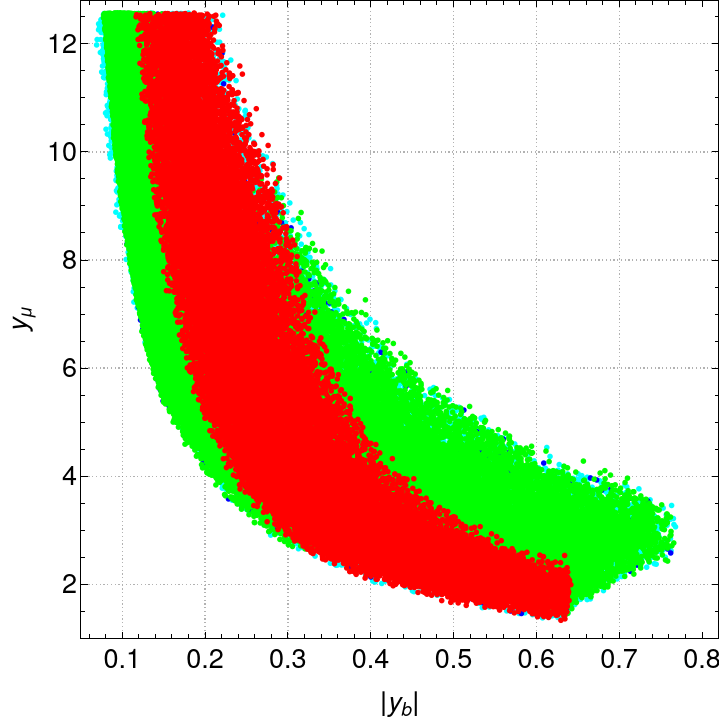}
		\includegraphics[height = 6.cm]{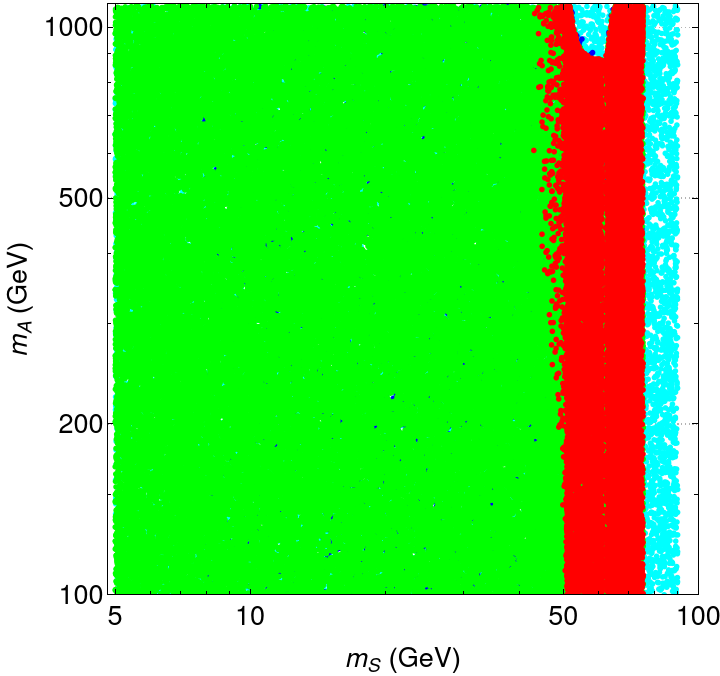}
		\\
		\vspace{2mm}
		\includegraphics[height = 6.cm]{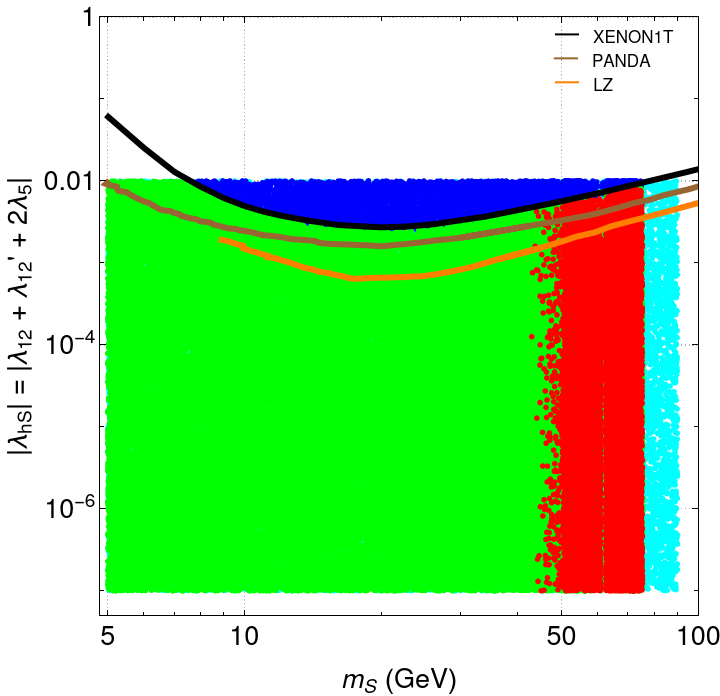}
		\includegraphics[height = 6.cm]{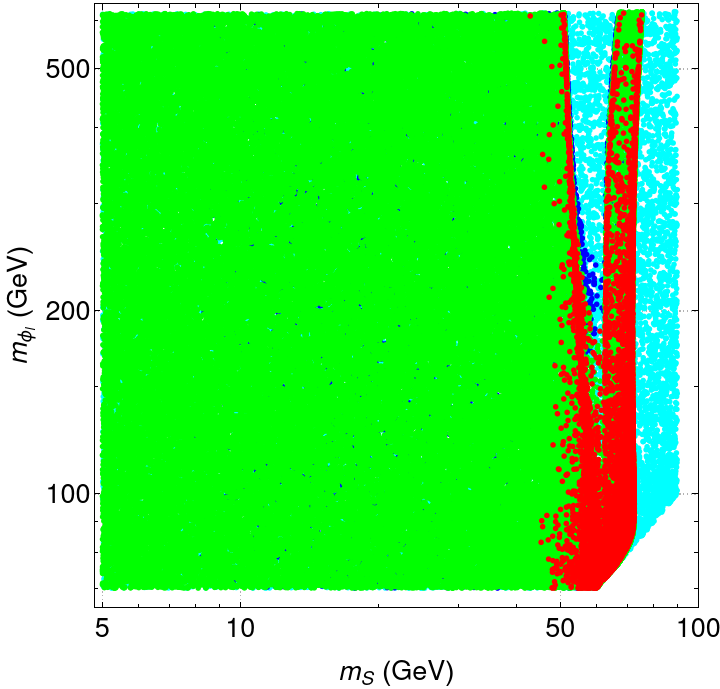}
		\caption{Model 3 allowed parameter space in the $(m_\chi, |y_b|)$ (top left), $(m_\chi, y_\mu)$ (top right), $(|y_b|, y_\mu)$ (middle left), $(m_S, m_A)$ (middle right), $(m_S, |\lambda_{hS}|)$ (bottom left) and $(m_S, m_{\phi_l})$ (bottom right) planes. The values used for the parameters are the ones from scan II. The solid black, brown and orange lines represent the experimental upper bounds provided by XENON1T, PandaX-4T and LZ, respectively. }
		\label{fig:res1}
	\end{center}
\end{figure}  
The main results for Model 3, which were obtained using scan II, are shown in Figure~\ref{fig:res1}. In the first three plots of Figure~\ref{fig:res1}, we see that we have sizeable Yukawa couplings with similar limits as the ones in Model 5, as expected since the flavor physics in both models is the same and the DM constraints do not have a major impact on the parameter space shown in these plots, with $y_\mu > 1.3$ and $0.11 < |y_b| < 0.65$ when all constraints are taken into account. 
In the next three plots of the same figure, we show the data points of our model as a function of the variables most relevant to the DM physics. As we had already seen in Figure~\ref{fig:DMrelic}, the DM relic density limits in a significant way the allowed values for the DM mass, imposing $m_S < 80$~GeV. By further taking into account the $g-2$ constraint, 42~GeV $< m_S <$ 76~GeV (in Model 5, 30~GeV $< m_S <$ 350~GeV). This is a significant distinction in the allowed parameter space of both models: the DM mass is limited in a very narrow range in Model 3, while for Model 5 its range is much broader. For the remaining parameters we observe $m_A < 1076$~GeV (middle right plot), the bounds on $|\lambda_{hS}|$ from XENON1T, PandaX-4T and LZ are in the bottom left plot, and $m_{\phi_l} < 621$~GeV (bottom right plot). The lower limit on $m_{\phi_l}$ was expected, as this is necessary to keep $\lambda_{12} < 4\pi$.

\begin{figure}[!ht]
	\begin{center}
	\hspace{-5mm}
		\includegraphics[height = 6.cm]{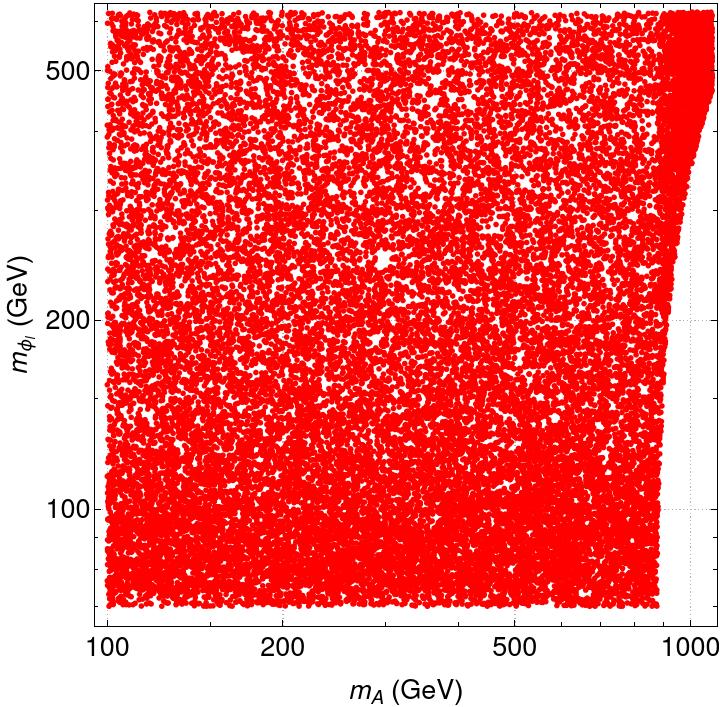}
		\hspace{5mm}
		\includegraphics[height = 6.3cm]{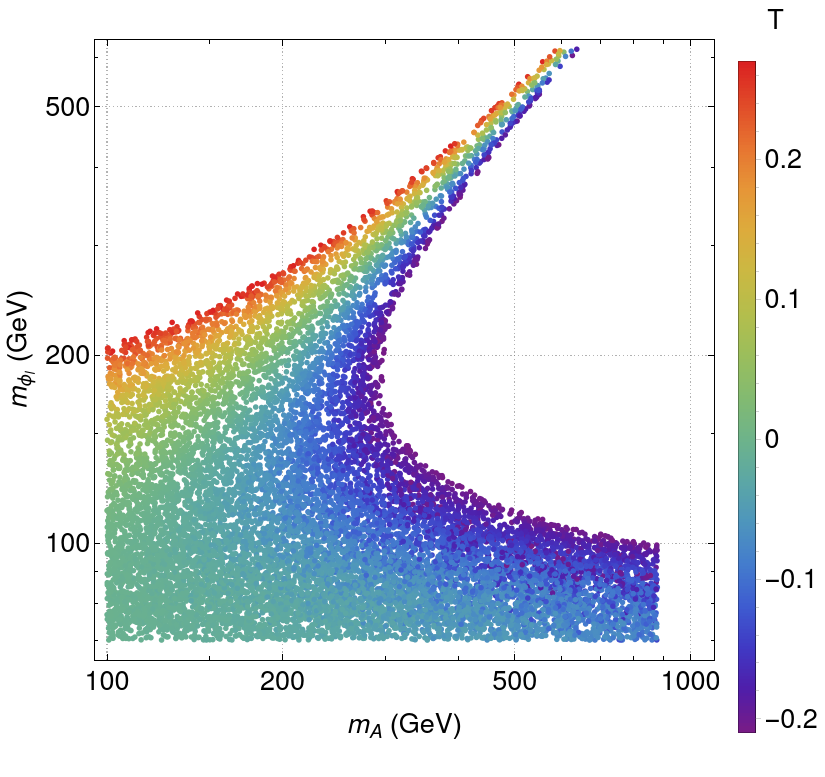}
		\caption{Model 3 allowed parameter space in the $(m_A, m_{\phi_l})$ plane, before applying the limits for the oblique parameter $T$ (left), and after, as a function of the values of $T$ (right). Only the points which verify all the previously mentioned constraints are shown.}
		\label{fig:Tparam}
	\end{center}
\end{figure}

\begin{figure}[!th]
	\begin{center}
		\hspace{-5mm}
		\includegraphics[height = 6.2cm]{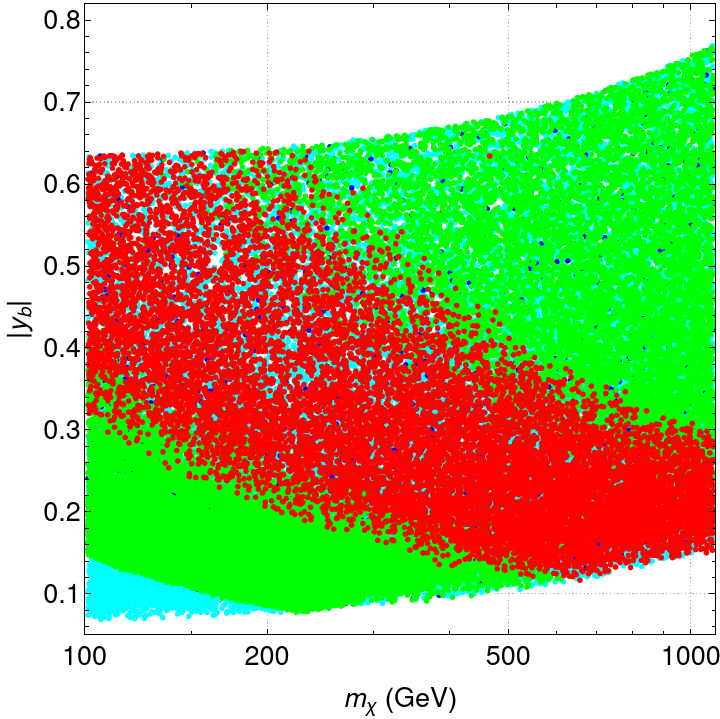}
		\hspace{2mm}
		\includegraphics[height = 6.2cm]{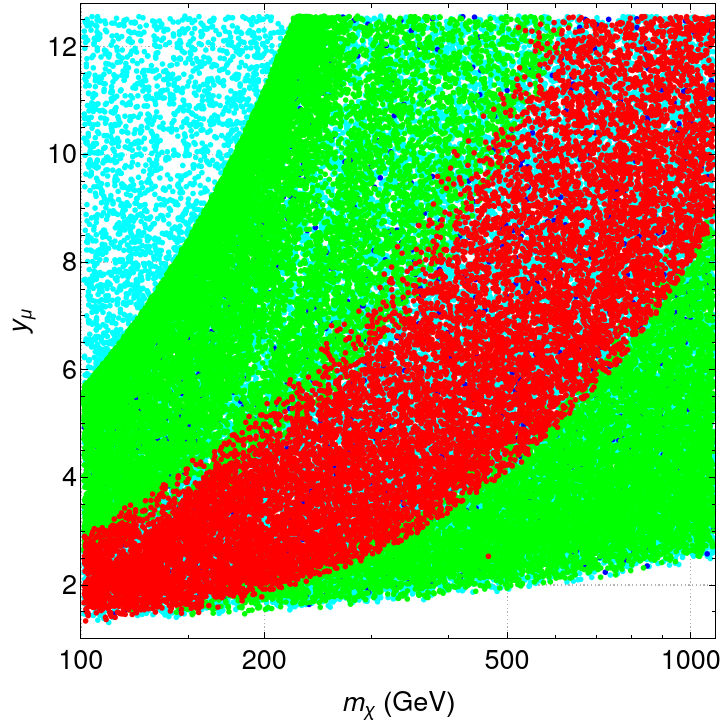}
		\\
		\vspace{2mm}
		\includegraphics[height = 6.2cm]{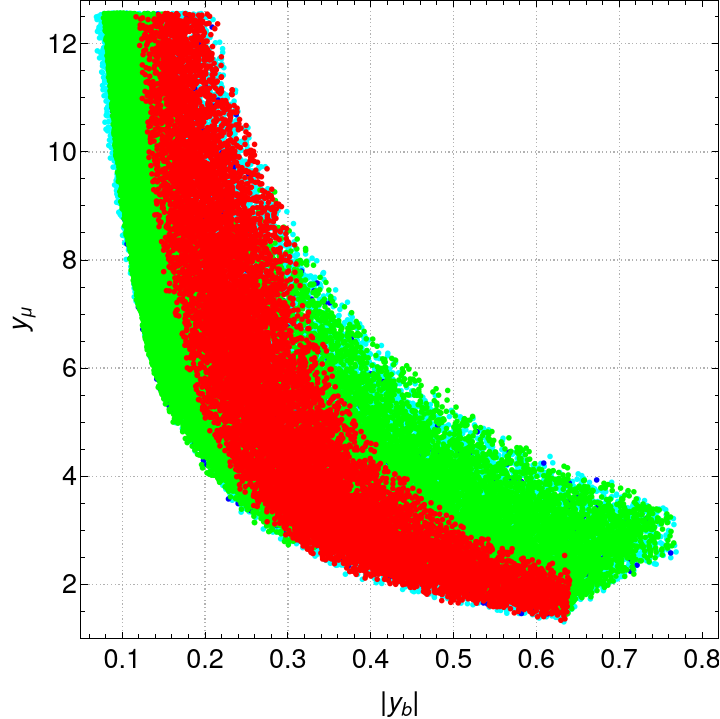}
		\includegraphics[height = 6.2cm]{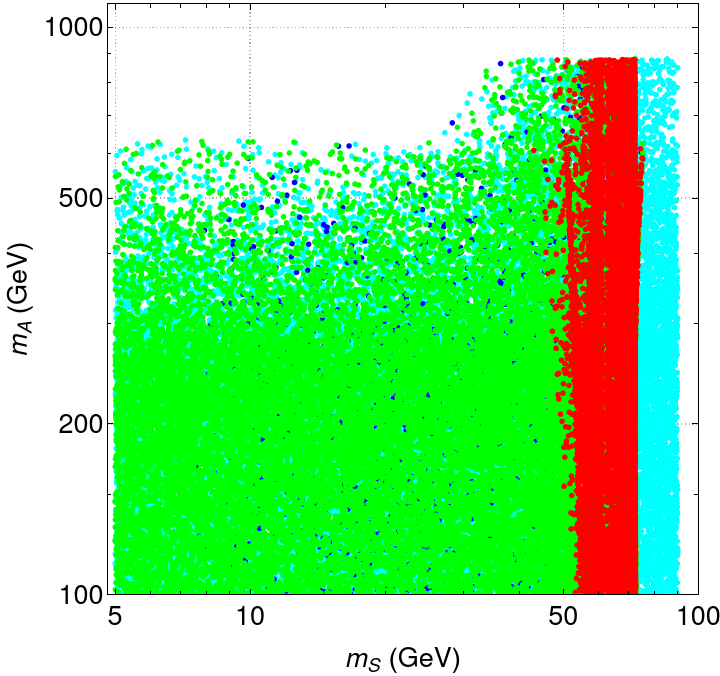}
		\\
		\vspace{2mm}
		\includegraphics[height = 6.2cm]{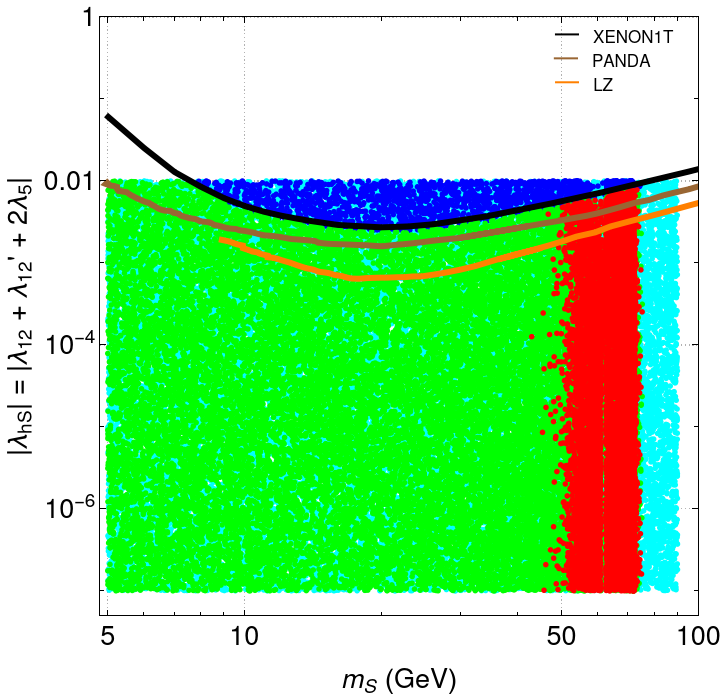}
		\includegraphics[height = 6.2cm]{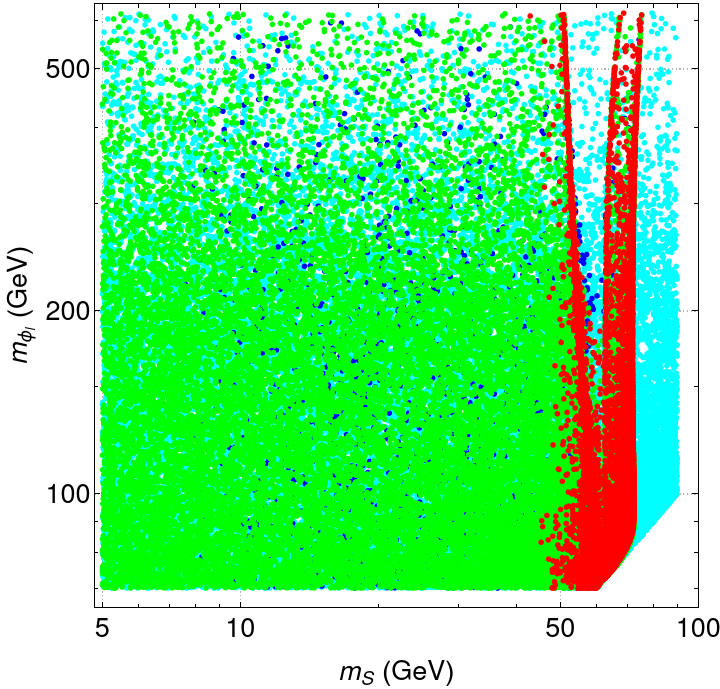}
		\caption{Model 3 allowed parameter space in the $(m_\chi, |y_b|)$ (top left), $(m_\chi, y_\mu)$ (top right), $(|y_b|, y_\mu)$ (middle left), $(m_S, m_A)$ (middle right), $(m_S, |\lambda_{hS}|)$ (bottom left) and $(m_S, m_{\phi_l})$ (bottom right) planes. All points verify the $2\sigma$ experimental bounds for the oblique parameter $T$. The solid black, brown and orange lines represent the experimental upper bounds provided by XENON1T, PandaX-4T and LZ, respectively. }
		\label{fig:res2}
	\end{center}
\end{figure}

Finally, after applying the limits for the oblique parameter $T$ to the allowed parameter space of Model 3 (see the end of Section~\ref{sec:2}), we observe two main differences: the upper limit for the pseudoscalar Higgs mass goes down, from $m_A < 1076$~GeV to $m_A < 877$~GeV, and for heavier masses ($m_{\phi_{l}} > 200$~GeV and $m_A > 300$~GeV), the vast majority of the previously allowed parameter space is now excluded. This is shown in Figure~\ref{fig:Tparam}. The points shown are the ones that verify all previous constraints. The effect of the $T$ parameter is to preferably select regions where $m_{\phi_{l}} - m_A \approx 0$, since this leads to $T \approx 0$. This is particularly true for $m_{\phi_{l}} > 200$~GeV, as we can see on the right side of Figure~\ref{fig:Tparam}. Nevertheless, because we can make the approximation $T \propto (m_{\phi_{l}} - m_A)(m_{\phi_{l}} - m_S)$~\cite{Barbieri_2006}, significant mass splits can still exist for small values of $m_{\phi_{l}}$, where $m_{\phi_{l}} \approx m_S$. For larger values of $m_{\phi_{l}}$, the only way to keep $T$ in its experimental bounds is to have $m_{\phi_{l}} \approx m_A$, which is why a significant part of the parameter space is excluded in this region. Although only the $T$ variable was used, the $S$ variable is not expected to be as sensitive to mass splits, since it depends on the scalar particle masses only logarithmically~\cite{Barbieri_2006, Belyaev_2018}. For completeness, we show in Figure~\ref{fig:res2} the main results of this paper again, but now all the points are within the $2\sigma$ experimental bounds for the oblique parameter $T$.

\begin{figure}[!ht]
	\begin{center}
	\hspace{-5mm}
		\includegraphics[height = 6.cm]{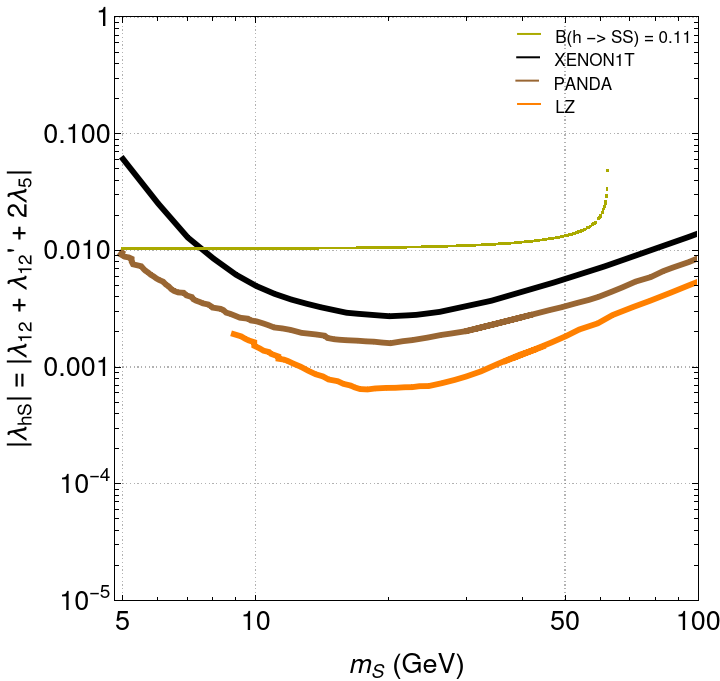}
		\caption{Comparison of DM direct detection experiments with the measurement of the Higgs invisible width, with the portal coupling shown as a function of the DM mass.}
		\label{fig:DDvs}
	\end{center}
\end{figure}

We end this section with a comparison between direct detections and collider bounds for future experiments. First, remember that the two observables are proportional to exactly the same portal coupling. Therefore
the constraints obtained are on the portal coupling as a function of the DM mass. In Figure~\ref{fig:DDvs} we show the most recent DM direct detection bounds together with the latest LHC measurement of the Higgs invisible width. There is already more than one order of magnitude difference between the LZ experiment and the LHC measurement. Therefore it is not expected
that future measurements of the Higgs invisible width would be able to compete with direct detection experiments.

\section{Conclusions \label{sec:conclusions}}
\hspace{\parindent} 
\label{sec:6}

We have explored a model belonging to a class that provides a solution to the lepton flavor universality violation observed in $b\to s\mu^+ \mu^-$ by the LHCb and Belle Collaborations. 
The model also provides a DM candidate and solves  the muon $(g-2)$ anomaly. In a previous work~\cite{PhysRevD.102.075009} a model was discussed where the main difference with the present  
work was in the group representation of the new scalars and of the new fermion fields, from the dark sector. In the previous model, Model 5, we have introduced an $SU(2)_L$ doublet vectorlike fermion $\chi$ and two complex scalar singlets, 
$\Phi_q$ and $\Phi_l$, the former is an $SU(3)_c$ triplet while the latter is colourless. The model discussed here, Model 3, is built such that  the vectorlike fermion $\chi$ is a singlet, while
 $\Phi_q$ and $\Phi_l$ are $SU(2)_L$ doublets. Here $\Phi_q$ is still an $SU(3)_c$ triplet. We have thoroughly studied the flavor and DM phenomenology in the two models.

The question we wanted to answer was how different group representations affected the allowed parameter space of the models. First of all the structure of the new Yukawa Lagrangian is
such that the actual vertices contributing to the loop processes are the same. This in turn means that both the contributions to flavor observables and to the muon $g-2$ do not change. 
There are however two crucial differences in what concerns the DM observables. First, in order to comply with the experimentally observed relic density by the Planck Collaboration
the DM mass has to be below about 80 GeV in Model 3 while in the previously studied Model 5 such restriction did not exist. This difference is related to the different 
group representations. In fact since in Model 3 the scalar fields couple to gauge bosons, the very efficient annihilation processes $SS \to W^+ W^-$ and $SS \to ZZ$ lead to a very
small relic density contribution, similarly to what happens in the Inert doublet model. Second, there is a striking difference in the Higgs portal coupling. Again, due to the group representation, while in Model 5 the portal coupling is a free input parameter, in Model 3 it is constrained and is defined as $\lambda_{hS} = \lambda_{12} + 2 \, (m_S^2 - m_{\phi_{l}}^2)/v^2$. 
A small portal coupling can be attained by choosing  $\lambda_{12} $ and $(m_S^2 - m_{\phi_{l}}^2)/v^2$ simultaneously small or $\lambda_{12} \approx - 2 \, (m_S^2 - m_{\phi_{l}}^2)/v^2$.
Owing to the bound coming from the relic density measurement the latter is the only viable option. In practice we have varied the portal coupling from $10^{-7}$ to $10^{-2}$.

In conclusion, the DM constraints act on the two models in a dramatically different manner. This difference is translated into distinct allowed mass regions for the DM particle.
While in Model 5 the allowed range is 30~GeV $< m_S <$ 350~GeV, in Model 3 this range is reduced to  42~GeV $< m_S <$ 76~GeV. Over the last year we have seen two
new released bounds on the DM direct detection from PandaX-4T and from LZ~\cite{LZ:2021xov, LZnew}. It is clear that these bounds have decreased the allowed value
of the portal coupling but the allowed mass region did not change.

\vspace*{1cm}
\subsubsection*{Acknowledgments}

We thank Jo\~ao Paulo Silva for discussions. RC and RS are partially supported by the Portuguese Foundation for Science and Technology (FCT) under Contracts no. UIDB/00618/2020, UIDP/00618/2020, PTDC/FIS-PAR/31000/2017 and CERN/FIS-PAR/0014/2019. RC is additionally supported by FCT grant 2020.08221.BD. {DH is supported in part by the National Natural Science Foundation of China (NSFC) under Grant No. 12005254, the National Key Research and Development Program of China under Grant No. 2021YFC2203003, and the Key Research Program of Chinese Academy of Sciences under grant No. XDPB15.} TL is partially supported by CFTP-FCT Unit 777
(UIDB/00777/2020 and UIDP/00777/2020), PTDC/FIS-PAR/29436/2017, CERN/FIS-PAR/0008/2019 and CERN/FIS-PAR/0002/2021, which are partially funded through POCTI (FEDER), COMPETE, QREN and EU.

\vspace*{1cm}
\bibliographystyle{h-physrev}
\bibliography{model3.bib}

\end{document}